\documentclass[twocolumn,showpacs]{revtex4}

\usepackage{amsmath}
\usepackage{amssymb}
\usepackage{graphicx}
\usepackage{wrapfig}
\usepackage{color}
\usepackage{bm}

\newcommand{\mem}{{\mathrm{m}}}
\newcommand{\sigm}{{\sigma}}
\newcommand{\shear}{\mathrm{shear}}

\newcommand{\mc}{\mathcal}
\newcommand{\be}{\begin{equation}}
\newcommand{\ee}{\end{equation}}
\newcommand{\bea}{\begin{eqnarray}}
\newcommand{\eea}{\end{eqnarray}}
\newcommand{\bs}{\boldsymbol}

\newcommand{\mbf}{\mathbf}
\newcommand{\im}{\mathrm{i}}

\newcommand{\bt}{\mathbf{t}}
\newcommand{\bn}{\mathbf{n}}
\newcommand{\out}{{\mathrm{ex}}}
\newcommand{\ins}{{\mathrm{in}}}
\newcommand{\hd}{{\mathrm{hd}}}
\newcommand{\el}{{\mathrm{el}}}

\newcommand{\bv}{{\bf v}}
\newcommand{\bu}{{\bf u}}

\newcommand{\bzh}{{\mathbf{\hat{z}}}}
\newcommand{\byh}{{\mathbf{\hat{y}}}}
\newcommand{\bxh}{{\mathbf{\hat{x}}}}

\newcommand{\refeq}[1]{(\ref{#1})}
\newcommand{\bE}{{\bf E}}
\newcommand{\bT}{{\bf T}}

\newcommand{\Ca}{\mbox{\it Ca}}
\newcommand{\Mn}{\mbox{\it Mn\,}}
\newcommand{\G} {{\dot\gamma}}
\newcommand{\Sr}{{S}}
\newcommand{\Rr}{{\Lambda}}
\newcommand{\visrat}{{\eta}}
\newcommand{\ext}{{\mathrm  e}}
\newcommand{\ehd}{{\mathrm  {el}}}
\newcommand{\cp}{{\mathrm  {m}}}

\newcommand{\half} {{\frac{1}{2}}}
\newcommand{\rhat}{{\bf{\hat r}}}
\newcommand{\bS}{{\bf y}}
\newcommand{\br}{{\bf r}}

\newcommand{\brac}[1]{\left(#1\right)}

\newcommand{\bnabla}{\nabla}

\begin{document}

\title{Vesicle electrohydrodynamics}

\author{Jonathan T. Schwalbe$^1$, Petia M. Vlahovska$^2$ and Michael J. Miksis$^1$}
\affiliation{$^1$Department of Engineering Sciences and Applied Mathematics, Northwestern University, Evanston, IL 60202, USA\\$^2$School of Engineering, Brown University, Providence, RI 02912, USA}

\date{\today}

\begin{abstract}

A small amplitude perturbation analysis is developed to describe the effect of a
uniform electric field on the dynamics of a lipid bilayer vesicle in a simple shear flow. All media are treated as leaky dielectrics and fluid motion is described by the Stokes equations. The instantaneous vesicle shape is obtained by balancing electric,
hydrodynamic, bending, and tension stresses exerted on the membrane.  We find  that in the absence of ambient shear flow, it is possible that an applied step--wise uniform DC electric field could cause the vesicle shape to evolve from oblate to prolate over time if the encapsulated fluid is less conducting than the suspending fluid. For a vesicle in ambient shear flow, the electric field damps the tumbling motion leading to a stable tank-treading state.


\end{abstract}

\maketitle

\section{Introduction}

 Membranes that encapsulate cells and internal cellular organelles are composed primarily of lipid bilayers \cite{Alberts}. Giant unilamellar vesicles (GUVs), which are cell-size membrane envelopes, have gained popularity as models of protocells \cite{Walde:2010}  and  systems to study membrane biophysics \cite{Rumy:2006}.  Because their large size (10-100~$\mu m$), direct observation is possible of the dynamic features of individual membrane vesicles in real time with optical microscopy. GUVs exhibit rich dynamic behavior in flow or electric fields, see for example the reviews \cite{Abkarian:2008, VlahovskaCR, softmatter:2009, VlahovskaAPLB}.  Understanding the effects of flow on GUV's and cells is fundamental to many naturally occurring biological processes, e.g., blood flow.  Applied electric fields are of recent interest because of the possible applications to biomedical technologies, e.g., gene transfection. In particular, a controlled application of an electric pulse can induce transient pores in the cell or vesicle membrane, which can reseal after the pulse is turned off but may allow the delivery of exogenous molecules.  Here we will investigate the combined effect of both flow and an applied DC electric field on the dynamics of a vesicle.

In simple shear flow,  a vesicle exhibits   several different types of motions.  A key physical parameter affecting the dynamics is the viscosity ratio between the fluid outside to the fluid inside the vesicle.  With varying viscosity ratio, three  of the observed dynamics are \cite{Kantsler-Steinberg:2005, Kantsler-Steinberg:2006, Mader:2006, Kantsler-Steinberg:2009, Deschamps:2009}: (1)~{\it tank--treading} (TT), in which  the vesicle  deforms into a prolate ellipsoid and the membrane  rotates as a tank-tread, the vesicle major axis is tilted with respect to the flow direction and the inclination angle remains fixed in time;
 (2) {\it tumbling} (TB), in which the vesicle undergoes a periodic flipping motion; and (3) {\it vacillating-breathing} (VB)  also called {\it trembling}, where the vesicle is trembling in the flow direction with periodic shape deformations.

A vesicle  deforms into an ellipsoid when subjected to a uniform electric field \cite{Kummrow-Helfrich:1991, Mitov:1993, Riske-Dimova:2005, Aranda:2008, Antonova-Vitkova:2010}.  Depending on the conductivity mismatch between the inner and outer fluids, and in the case of the AC field, its frequency, the ellipsoid is prolate or oblate and its major axis is either collinear with or perpendicular to the applied electric field \cite{Aranda:2008, Riske-Dimova:2006}.

If a simple  shear flow and electric field are simultaneously applied, vesicle deformation and  orientation become dependent on the relative strength of the electric and shear stresses. For example, an electric field applied along the velocity gradient acts to elongate and align the vesicle perpendicularly to the flow direction, while the shear flow tends to orient the vesicle  along the flow direction. This problem, however, has been analyzed only to a limited extent for  drops and capsules \cite{Allan-Mason:1962, Ha:2000c,  Papageorgiou:2009,Vlahovska:ER}.

The goal of this paper is to theoretically investigate the effect of the competition between electric  stress and shear stress on vesicle dynamics.  While the behavior of an isolated vesicle in either uniform electric field  \cite{Vlahovska-Dimova:2009} or linear flows \cite{Misbah:2006, Vlahovska:2007, Lebedev:2008, Schwalbe:2010} has been extensively studied, the effect of a combined uniform electric and and fluid flow on vesicle dynamics has received no attention. Our study is also motivated by the possible use of electric fields to modulate rheology of vesicle suspensions, and in more general context to use electrohydrodynamics  for cell manipulation.

The theoretical analysis of a vesicles in external flows is complicated by the elasto- and electromechanics of the lipid bilayer membrane. Several features  of lipid membranes can be identified which underly the complexity of the problem: (1) {Lipid molecules are free to move  in the plane of the membrane thus the  lipid bilayer behaves as a fluid;} (2) {Under stress,  lipid bilayers store elastic energy in bending, while membranes made of cross-linked polymers are more likely to be stretched and sheared;} (3){The lipid bilayer contains a fixed number of molecules and the  membrane is nearly area--incompressible. In response to in--plane stresses, it develops nonuniform tension, which adapts itself to the  forces exerted on the membrane in order to keep the local area constant;} and (4){ The lipid membrane is essentially an insulating shell impermeable to ions. When an electric field is applied,  charges accumulate on both sides of the bilayer and the  membrane acts as a charging capacitor.} In addition, since membranes are embedded in a fluid environment, changes in membrane conformation are coupled to a motion in the surrounding fluids.

Since membranes are molecularly thin, to describe the membrane-fluid coupling  it is convenient to use an effective two--dimensional description of the membrane mechanics \cite{Lomholt-Miao:2006}.  The simplest account for the bending stresses comes  from the classic Helfrich--Canham energy \cite{Helfrich:1973, Canham:1970}.
In this paper we develop an effective  zero--thickness model for a fluid-embedded lipid membrane in an electric field and apply it  to study vesicle dynamics in a combined shear flow and uniform electric fields.

\section{Problem formulation}
\label{electric}
\subsection{The physical picture: characteristic time scales, relevant parameters, and their magnitudes}
Let us consider a  neutrally--buoyant vesicle made of a
charge-free lipid bilayer  membrane with  conductivity $\sigm_{\mem}$ and dielectric constant $\epsilon_\mem$. The bilayer thickness is about $h\sim~5nm$, thus  on the length scale of a cell-size vesicle (radius $a\sim~10\mu m$) the bilayer membrane can be regarded as a two-dimensional surface with capacitance $C_\mem=\epsilon_\mem/h$ and conductivity $G_\mem=\sigm_{\mem}/h$.
The vesicle is filled with a fluid of viscosity
$ \mu_\ins$, conductivity $\sigm_{\ins}$, and dielectric constant $\epsilon_\ins$, and suspended in a different
fluid characterized by $ \mu_\out$,  $\sigm_{\out}$, and  $\epsilon_\out$.
 To characterize the mismatch in the fluid physical properties, we introduce the ratios
\begin{equation}
\Rr=\frac{\sigm_\ins}{\sigm_\out}\,,\quad \Sr=\frac{\epsilon_\ins}{\epsilon_\out}\,,\quad \visrat=\frac{ \mu_\ins}{ \mu_\out}\,.
\end{equation}

The departure of the vesicle shape from a sphere is quantified by the excess area, which is the difference between the vesicle area and the area of an equivalent-volume sphere \cite{Seifert:1997}
\begin{equation}
\label{Delta}
\Delta=A/a^2-4\pi \,,\quad a=\left(\frac{3v}{4\pi}\right)^{1/3}\,.
\end{equation}
Here $A$ and $v$ are the true surface area and volume of the vesicle.

\begin{figure}[h]
\centerline{\includegraphics[height=1.25in]{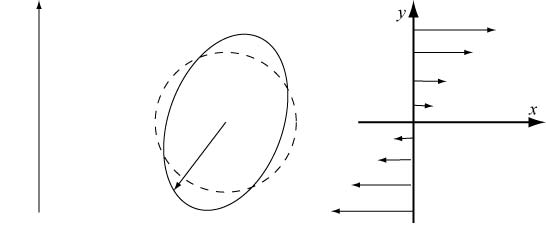}}
\begin{picture}(0,0)(0,0)
\put(60,100){$\bu^\infty=\G y  \bxh$}
\put(90,60){$ \phi$}
\put(80,55){\rotatebox{120}{$\rightarrow$}}
\put(-100,100){{$\bE^\infty=E_0  \byh$}}
\put(-40,55){$ \mu_\ins,\,\epsilon_\ins,\, \sigm_\ins$}
\put(-20,90){$ \mu_\out,\,\epsilon_\out,\, \sigm_\out$}
\put(-50,5){{${r_s=1+ f(\theta, \phi,t)}$}}
\end{picture}
\caption{\footnotesize Sketch of the the problem: a vesicle subjected to a combination of shear flow and a uniform electric field.}
\label{figsketch}
\end{figure}
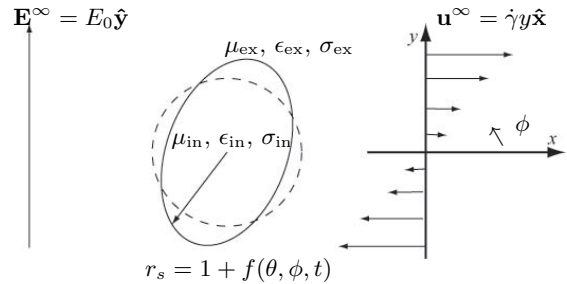

The vesicle is subjected to  a linear flow with strain-rate magnitude
$\G$ and a uniform DC electric field with magnitude $E_0$,
\begin{equation}
\bu^\infty=\G y \bxh\,, \qquad \bE^\infty=E_0\byh\,.
\end{equation}
The vesicle shape can be described by the radial position of the interface  $r_s =a(1 + f(\theta, \phi,t))$, where $f(\theta, \phi,t)$ is to be determined  as part of the solution.
The problem is sketched in Figure \ref{figsketch}.
\vspace{0.25cm}

\begin{figure}[h]
 \centerline{\includegraphics[height=0.9in]{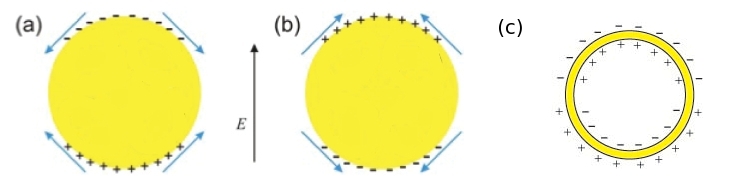}}
     \caption{\footnotesize Surface  charge distribution and direction of the surface electric force for a sphere with $t_{c,\ins}>t_{c,\out}$ (a) and $t_{c,\ins}<t_{c,\out}$ (b). (c) Sketch of the induced charge distribution around a spherical insulating shell.}
      \label{figT}
\end{figure}

When an electric field $E(t)$ is applied to an electrolyte solution,
 ions move.  The  ion redistribution leads to
inhomogeneities in the bulk charge density, which  decay on a  time scale  related to bulk conduction \cite{Melcher-Taylor:1969, Saville:1997}
\begin{equation}
\label{tc}
t_{c,\ins}=\frac{\epsilon_\ins}{\sigm_\ins}\,, \qquad t_{c,\out}=\frac{\epsilon_\out}{\sigm_\out}\,.
\end{equation}
Free charges accumulate  at
 boundaries that separate media with different electric properties as illustrated in Figure \ref{figT}. The rate of  charge build-up at the  interface of a macroscopic object, e.g., a sphere, is given  by the Maxwell-Wagner polarization time \cite{JonesTB}
\begin{equation}
\label{tMW}
t_{MW}=\frac{\epsilon_\ins+2\epsilon_\out}{\sigm_\ins+2\sigm_\out}\,.
\end{equation}
The polarization depends on $t_{c,\ins}/t_{c,\out}=\Lambda/S$. Consider, for example, a droplet suspended in another liquid. The charge relaxation time, $t_c$, measures how  fast conduction supplies charges to restore equilibrium.
If $t_{c,\ins}<t_{c,\out}$, the conduction in the drop is faster than  the suspending liquid.  As a result, the interface acquires charge  dominated by ions brought from the interior fluid and the induced dipole is aligned with the electric field.  In this case, charges at the poles are attracted by  the electrodes, pulling the drop into a prolate shape.
In the opposite case, $t_{c,\ins}>t_{c,\out}$, the charging response of the exterior fluid is faster than the interior fluid.  Hence, the interface charge is dominated by the exterior  ions and the polarization is reversed.
In this induced--charge configuration, a drop can become  an oblate ellipsoid \cite{Taylor:1966}.
The lipid membrane, however, represents a more complex boundary compared to fluid-fluid interfaces.
It is impermeable to ions and, therefore, charges accumulate on both the inner and outer physical surfaces. Hence,
the  vesicle acts as a  capacitor that charges on a time scale given by  \cite{Grosse-Schwan:1992, Schwan, Kinosita:1988}
\begin{equation}
\label{tcapV}
t_{\cp}=a C_m \left(\frac{1}{\sigm_\ins}+\frac{1}{2\sigm_\out}\right)\,.
\end{equation}
For simplicity,  the vesicle is modeled as a spherical insulating shell. The membrane capacitance gives rise to a potential difference across the membrane and a capacitive current through the membrane.

If the electric field is not normal to the interface, its tangential component  acts on the induced  free charges at the interface  and  gives rise to a  shearing  force. This is illustrated in Figure \ref{figT} on the example of a spherical droplet. The electrical force drags the interface in motion. The resulting electrohydrodynamic (EHD) flow is characterized by a time scale, which  corresponds to the inverse of the shear rate imposed by the tangential electric stress
\begin{equation}
t_{\ehd}=\frac{  \mu_\out (1+\visrat)}{\epsilon_\out E_0^2}\,.
\end{equation}

The straining component of the external shear flow also distorts vesicle shape by elongating
it  along the extensional axis of the flow, which is oriented at $45^o$ angle relative to the flow direction. The corresponding time scale is $t_{\ext}=(1+\visrat)\G^{-1}$.

Vesicle  deformation by electric and flow stresses is limited by the membrane's resistance to bending and stretching. A distortion in the vesicle shape  relaxes on a  timescale
\begin{equation}
t_{\kappa}=\frac{ \mu_\out (1+\visrat) a^3}{\kappa}\,,
\end{equation}
where $\kappa$ is the bending modulus. The curvature relaxation depends on the average viscosities of the bulk fluids, because viscous dissipation on lengthscales greater than a micrometer takes place in the bulk~\cite{Seifert:1999}.

The ratio of distorting electric and restoring bending time scales defines a capillary--like number
\begin{equation}
\label{capillary number}
\Ca =\frac{t_{\ehd}}{t_\kappa}\equiv\frac{  \epsilon_\out E_0^2 a^3}{\kappa}
\end{equation}
It is convenient  to introduce a dimensionless number, which is independent of the membrane properties
\begin{equation}
\label{Mason number}
\Mn=\frac{ \epsilon_\out E_0^2}{ \mu_\out \G} \,.
\end{equation}
The Mason parameter, $\Mn$,  compares the strength of  electric and viscous stresses.

Let us estimate the magnitude of the above time scales involved in the process of vesicle electrodeformation. Typical experimental conditions involve solutions with conductivities in the range  $\sigm\sim10^{-3}-10^{-4} S/m$ and electric fields of the order of  $E\sim 1kV/cm$ \cite{Kummrow-Helfrich:1991, Mitov:1993, Riske-Dimova:2005, Riske-Dimova:2006, Dimova-Aranda:2007, Aranda:2008, Portet:2009, Riske-Dimova:2009, Antonova-Vitkova:2010, Gracia:2010, Portet:2010,Knorr:2010}.  The typical size of a giant vesicle  is $a\sim 10\mu m$. The inner and outer fluids are essentially water: viscosity $ \mu\sim10^{-3} Pa.s$, and density  $\rho\sim1000 kg/m^3$. The membrane capacitance is $C_m \sim 10^{-2} F/m^2$ \cite{Needham-Hochmuth:1989} and bending rigidity $\kappa\sim10^{-19} J$. Therefore, for vesicles, we estimate the basic charging time and the Maxwell-Wagner polarization time to be of the same order  $t_c\sim t_{MW} \sim 10^{-7}s$, the membrane charging time  is $t_{\cp} \sim 10^{-3} s$,  the electrohydrodynamic time is $t_\el \sim 10^{-3} s$, and the bending relaxation time is $t_\kappa \sim 10 s$. Typical shear rates range from $\G\sim0.1\, s^{-1}$ to $100\,s^{-1}$ \cite{Kantsler-Segre-Steinberg:2008b,Vitkova:2008, Coupier:2008,Callens:2008}.

We see that the vesicle electrohydrodynamics involves processes that occur on very different timescales. Bulk phases become electro-neutral on a time scale given by charge relaxation time \refeq{tc}, and charging of the interface occurs on a similarly fast time-scale \refeq{tMW}. Hence, we can assume a quasi--static electric field. However, the electric field can vary with time as the membrane capacitor charges. These variations can take place on a time-scale comparable  to vesicle response to imposed shear flow ($t_{\cp} \sim \G^{-1}$ ), or electric field ($t_{\cp} \sim t_\ehd$) rendering the problem intrinsically non-linear and time--dependent.

Henceforth, all  quantities are rescaled using $\epsilon_\out$, $\sigm_\out$, $ \mu_\out$, $a$,  $\G$, and $E_0$.  For the electrostatic problem (in  absence of ambient shear flow), the time  scale is chosen to be $t_\ehd$, the charge scale is $\epsilon_\out E_0$,  and  bulk electric stress are scaled with $\epsilon_\out E_0^2$ . In the presence of shear, the time scale is $\G^{-1}$, the velocity scale
is $a\G$, and  bulk hydrodynamic stresses are scaled with $ \mu_\out \G$. The scaling of the electric charge and stresses remains unchanged.

{\subsection{Governing equations}}

We adopt the leaky dielectric model, which combines the Stokes equations to describe fluid motion with conservation of current described by Ohm's law \cite{Saville:1997}. Under the assumption of charge-free fluids, the electric and hydrodynamic fields are decoupled in the bulk.

The pressure, $p$, and the fluid velocity, $\bu$, fields obey
\begin{equation}
 \hat\visrat\nabla^2\bu=\nabla p \,,\qquad\nabla \cdot \bu=0 \,,
\end{equation}
where $\hat\visrat=1$ in the suspending fluid and $\hat\visrat=\visrat$ in the vesicle.
In the absence of bilayer slip and membrane permeability, the velocity is
continuous across the interface. The shape evolution is determined from the kinematic condition that the interface moves with the normal component of the fluid velocity  $\bu_\ins(r_s)=\bu_\out(r_s)\equiv \bu_s$
\begin{equation}\label{interfaceevolution}
\frac{\partial f}{\partial t}=\bu_s\cdot \rhat-\bu_s\cdot \nabla f\,.
\end{equation}

The quasi-static electric field, $\bE$, in the absence of bulk charges is solenoidal and the electric potential, $\Phi$, satisfies
\begin{equation}
\label{Laplaceeq}
 \bE=-\nabla \Phi\,,\qquad \nabla^2 \Phi=0 \,.
 \end{equation}
 The potential  undergoes a jump across a capacitive interface
\begin{equation}
\label{potential}
\Phi_\ins-\Phi_\out= V_\mem\, .
\end{equation}
The transmembrane potential, $V_\mem$, is determined as part of the problem; in general, it is a complex function of the geometry, and  fluid and membrane physical properties. Far away from the vesicle, the velocity $\bu$  and electric fields $\bE$ tend to the unperturbed flow, $\bu\rightarrow \bu^{\infty}$ and electric field, $\bE\rightarrow\bE^\infty$, respectively.

The electric and flow fields are coupled through the boundary condition for stress balance and current conservation at the interface. The hydrodynamic and electric tractions are discontinuous and are  balanced by membrane forces
\begin{equation}\label{stressEq}
\bn \cdot\left[\left( \bT_\out-\bT_\ins\right)+\Mn \left(\bT^{\el}_\out-\bT^{\el}_\ins\right)\right]={\bm{\tau}}^\mem\, \quad \mbox{at}\quad r=r_s \,,
\end{equation}
where $\bn$ is the outward pointing normal vector.
The membrane stresses $\tau^\mem$ are discussed in Section \ref{memtrac}.
Here $T_{ij}=-p\delta_{ij}+\hat \visrat(\partial_j u_i+\partial_i u_j)$ is the bulk hydrodynamic stress
  and $\delta_{ij}$ is the Kronecker delta function. The electric stress is given by the Maxwell stress tensor $T^\el_{ij}=\hat S \left(E_iE_j-E_iE_i\delta_{ij}/2\right)$; $\hat S=1$ in the suspending fluid and $\hat S=\Sr$ in the vesicle.

The conservation of normal current requires
\begin{equation}
\label{current conserv}
\bn \cdot \bE_\out=\Rr \bn \cdot \bE_\ins+(t_{c,\out} \dot\gamma)\nabla_s\cdot\left(\bu_s Q\right) \,.
\end{equation}
Charge convection along the surface by fluid motion is reflected by $\nabla_s\cdot \left(\bu_s Q\right)$. The effective induced charge on the membrane is formally defined as a jump in the displacement fields across the interface
\begin{equation}
\label{Enormal balance}
Q=\bn\cdot \left(\bE_\out-\Sr\bE_\ins\right)\,.
 \end{equation}
 In our model of the membrane as a zero-thickness capacitive interface,  $Q$ is not  the charge of the capacitor; for a fully charged capacitor $Q=0$. $Q$ represents the difference between the  charge densities on the inner and outer physical surfaces of the membrane. This imbalance occurs because if bulk conductivities differ, charges at the physical surfaces of the membrane are supplied at different rates.
 In order to make analytical progress, we neglect surface charge convection, i.e., we assume that bulk conduction  is much faster then the imposed shear flow, $t_{c,\out}\ll \G^{-1}$.

Charging of a capacitive interface gives rise to a transient displacement current. Hence, \refeq{current conserv} becomes
\begin{align}\label{charge conservation}
	\delta_\mem \frac{d V_{\mem}}{d t} = \bn \cdot \bE_\out=\Rr \bn \cdot \bE_\ins\,,
\end{align}
where charge convection has been neglected and $\delta_m=t_\mem\G /(\textstyle{{1}/{2}}+\Lambda^{-1})$ is the dimensionless capacitance.\\

\subsection{Membrane forces}
\label{memtrac}

Fluid membranes made of lipid bilayers are governed by resistance to curvature changes. The membrane free energy is
 \be\label{FreeEnergy}
	\mc{F} = \int\left[\frac{\kappa}{2}(2H)^2 + \Sigma\right]dA,
\ee
where $\kappa$ is the bending modulus. The membrane tension, $\Sigma$, is a Lagrange multiplier that enforces the area--incompressibility. The quantity $H$ is the mean curvature of the surface, given by
\be
	H = \frac{1}{2}\nabla\cdot\bn.
\ee
The corresponding membrane forces  are found by taking a variational derivative of \refeq{FreeEnergy} \cite[]{Seifert:1999}
\be
	{\bm{\tau}}^{\mem} = \left[-2\kappa\left(2H^3-2K_gH+\nabla_s^2 H\right)+2\Sigma H\right]\bn-\nabla_s\Sigma
\ee
where $K_g$ is the Gaussian curvature of the surface given by
\be
	K _g= \frac{1}{2}\nabla\cdot\left[\bn\nabla\cdot\bn+\bn\times(\nabla\times\bn) \right].
\ee
$\nabla_s = \mbf{I}_s\cdot\nabla$ is the surface gradient operator, $\mbf{I}_s = \mbf{I}-\bn\bn$ is the surface projection, and $I_{ij}\equiv \delta_{ij}$.

\section{Solution for a nearly spherical vesicle}

In order to make analytical progress, we consider a  vesicle with small excess area, $\Delta \ll 1$.  In this limit the deviation from sphericity, $f$, scales like $\Delta^{1/2}$.
We proceed to determine the leading--order solution.
Assuming that  the applied electric field scales as $\Delta^{1/4}$ allows us to find the electric field   by solving for the potential about a sphere with the boundary condition (\ref{current conserv}) independent of the flow and vesicle asphericity.  The corresponding electric stresses are then inserted in the stress conditions (\ref{stressEq}) to find the velocity field and the vesicle deformation.

\subsection{Solution outline}

Due to the linearity of the Stokes equations, the velocity field can be decomposed into two components:  a flow about a vesicle subject to a shear flow (in absence of electric field) and a flow about a vesicle in electric field (in absence of applied shear). The first problem has been solved in \cite{Misbah:2006, Vlahovska:2007}. Here we derive the solution for the second problem, namely, the electrodeformation of a  spherical particle with a capacitive interface. Then we combine the two solutions  and explore the vesicle dynamics resulting from the interplay of shear and electric stresses.  The solution of the hydrodynamic part is summarized in Appendix \ref{ap:solution}.

As noted earlier, in a spherical coordinate system centered at the vesicle, the position of the interface is
\begin{equation}\label{radialcoordinate}
 r_s(\theta,\phi,t)=1+  f(\theta,\phi,t)\,,
 \end{equation}
where $f$  measures the deviation from sphericity.
 All variables are expanded in spherical harmonics $Y_{jn}$ \refeq{normalized spherical harmonics}. For example,
\begin{align}\label{expansions}
	f(\theta,\phi,t) &= \sum_{j\geq2}\sum_{n=-j}^j f_{jn}(t)Y_{jn}\,.
\end{align}
The $j=1$ modes have been omitted because they describe translation of the center of mass. The order of magnitude of the asphericity ($f\sim \Delta^{1/2}$) becomes evident from the expression for the vesicle's excess area
\be\label{excessarea}
	\Delta = \frac{1}{2}\sum_{j,n}(j+2)(j-1)f_{jn}f_{jn}^*+O(f^3).
\ee
where the $\sum_{j,n}$ is shorthand notation for the double sum in \refeq{expansions} and the $*$ denotes the complex conjugate, $f_{jn}^* = (-1)^n f_{j-n}$.

The quasi--static electric field is irrotational, i.e.,  $\bE=-\nabla \Phi$,  and the electric potential $\Phi$ is a solution of the Laplace equation. Hence, the solutions for the electric field are growing and decaying spherical harmonics, which derive from $\nabla(r^jY_{jn})$ and $\nabla(r^{-j-1}Y_{jn})$
\begin{equation}
\begin{split}
\bE_\out=&\bE^\infty
-{\sum_{j,n}}P^\out_{jn}\nabla(r^{-j-1}Y_{jn})\,,\\
\bE_\ins=&-{\sum_{j,n}}P^\ins_{jn}\nabla(r^jY_{jn})\,
\end{split}
\end{equation}
A uniform electric field applied in the $y-z$ plane (perpendicular to the flow direction)
  is defined by
\begin{equation}
\label{Einf}
\begin{split}
\bE^\infty=\alpha\byh+\beta \bzh=-\sum_{n=-1}^{1}e^\infty_{1n}\nabla\left(r Y_{1n}\right)
\end{split}
\end{equation}
where
\begin{equation}
\label{coefs}
e^\infty_{10}=\beta \sqrt{\frac{4 \pi}{3}} \,,\qquad e^\infty_{1\pm1}= \alpha \im \sqrt{\frac{2 \pi}{3}}\,.
\end{equation}

\subsection{Electric field and  the transmembrane potential}
The  solution for the electric potential around a sphere placed in a uniform electric field
 is
\begin{equation}
\label{el field}
\begin{split}
\Phi_\out=&-[r + P_\out r^{-2}]\sum_{n=-1}^{1}e^\infty_{1n}Y_{1n}\,,\\
\Phi_\ins=&-P_\ins r\sum_{n=-1}^{1} e^\infty_{1n} Y_{1n}\,.
\end{split}
\end{equation}
Applying the boundary conditions \refeq{charge conservation}  we find
\begin{equation}
\label{el field2}
\begin{split}
P_\out= &\frac{(-\Rr+1)+\Rr \bar V(t)}{\Rr+2}\,,\\
P_\ins= & \frac{3-2\bar V(t)}{\Rr+2}\,,
\end{split}
\end{equation}
where $\bar V(t)$ is the amplitude of the transmembrane potential, $V_\mem=\Phi_\ins(r=1)-\Phi_\out(r=1)=\bar V(t) \sum e^\infty_{1n}  Y_{1n}$

\begin{equation}
\label{VmpotDC}
\bar V(t)=\frac{3}{2} \left[1-\exp\left(-\frac{t}{\delta_\mem(1/2+\Rr^{-1})}\right)\right] \,.
\end{equation}
Note that the transmembrane potential  is position dependent. Its absolute value is maximal at the poles, i.e. closest to the electrodes. At the equator the transmembrane potential is zero.
At steady state, the vesicle interior is ``shielded'', i.e., the interior electric field is zero, and the maximal potential drop across the membrane is $\bar V=1.5 $. The effective charge density is calculated from \refeq{Enormal balance}
\begin{equation}
\begin{split}
Q(t)&=\left(1-2 P_\out-\Sr P_\ins\right)\sum_{n=-1}^{1}e^\infty_{1n}Y_{1n}\\
&=\frac{\Rr-\Sr}{\Rr+2}\left(3-2\bar V(t)\right)\sum_{n=-1}^{1}e^\infty_{1n}Y_{1n} \,.
\end{split}
\end{equation}
We see that the effective charge on the membrane decreases as the transmembrane potential increase. At long times, when the capacitor becomes fully charged, the imbalance between the inner and outer surface charge density vanishes and $Q=0$.

\subsection{Electric stresses}
\label{ves:stress}
The tractions that the electric field exerts on a sphere are
\begin{equation}
\label{tE}
\begin{split}
\bt^\el=&\textstyle{\left[(\rhat\cdot\bE_\out)\bE_\out-\half \bE_\out\cdot\bE_\out\rhat\right]}\\
&\textstyle{-\Sr \left[(\rhat\cdot\bE_\ins)\bE_\ins-\half \bE_\ins\cdot\bE_\ins\rhat\right]}\,.
\end{split}
\end{equation}
A uniform electric field with $j=1$ symmetry generates electric tractions with  $j=0$ and $j=2$, see Appendix \ref{ap:harmonics}. The isotropic part $j=0$ is balanced by the hydrostatic pressure and does not lead to deformation. Only the position--dependent stress leads to vesicle deformation
\begin{equation}
\label{tel}
\bt^\el=\sum_{n=-2}^{2} \left(\tau^\el_{2n0} \bS_{2n0}+\tau^\el_{2n2}\bS_{2n2}\right)\,,
\end{equation}
where $\bS_{2n2}=Y_{2n}\rhat$ and $\bS_{2n0}$ are the vector spherical harmonics \refeq{vh2}. The electric tractions have a normal and tangential component
 \begin{equation}
 \label{el tractions}
\bt^\el=p^\el \rhat+{\bm{ \tau}}^s\,,
\end{equation}
The electric pressure is
\begin{equation}\label{pressure}
\begin{split}
p^\el=& - \bar p [3\alpha^2\cos 2\phi \sin^2\theta\\
& +\textstyle{\frac{1}{2}}(\alpha^2 -2 \beta^2)(1+3\cos 2\theta)].
\end{split}
\end{equation}
The tangential electric traction is
\begin{equation}
\label{taus}
\begin{split}
{\bm{ \tau}}^s=&\bar\tau^s\left\{-\alpha^2\sin\theta \sin2\phi\mathbf{e}_{\phi}\right.\\
&\left.+\textstyle{\frac{1}{2}}[\alpha^2(\cos2\phi-1)+2\beta^2]\sin 2\theta\mathbf{e}_\theta\right\}.
\end{split}
\end{equation}
We have explicitly shown the angular dependence of the pressure and tangential stress; the amplitudes $\bar p$ and $\bar \tau^s$ depend solely on the physical parameters of the system such as  $\Rr,\, \Sr$.
The amplitude of the radial (pressure) component is given by
\begin{equation}
\label{tn}
\bar p=\textstyle  \frac{1}{12}\left[2- 2 P_\out+5P_\out^2-2\Sr P_\ins ^2\right]\,,
\end{equation}
and the tangential (shearing) component is
\begin{equation}
\label{tt}
\bar \tau^s=\textstyle \frac{1}{2} \left[-1+ P_\out+2 P_\out ^2+\Sr P_\ins^2\right] \,.
\end{equation}

With the electrostatic problem solved, we next proceed to compute vesicle deformation in response to an electric field.

\section{Vesicle dynamics in absence of applied shear flow }\label{noflow}
At leading order (linear response), the vesicle shape has the same symmetry as the deformation--inducing electric stresses,  i.e. $j=2$. The  evolution equations for the  shape parameters $f_{2n}$ for a vesicle in an electric field oriented in an arbitrary direction are
 \cite{Vlahovska-Dimova:2009, VlahovskaAPLB, Schwalbe:thesis}
\begin{equation}
\label{ev eq}
 \frac{\partial f_{2n}}{\partial t}=C^\el_{2n}-\Ca^{-1} R_2f_{2n}\,.
\end{equation}
The inhomogeneous term
represents shape distortion by the applied electric field
\begin{equation}
\label{Celforcing}
	C^\el_{2n}=\frac{6\tau^\el_{2n2}+2\sqrt{6}\tau^\el_{2n0}}{32+23 \visrat}\,,
\end{equation}
$\tau^\el_{2nq}$ are defined in \refeq{tel}.  For a field in the $z$ direction
\begin{equation}
\begin{split}
	C^\el_{20}&=\frac{12 (3\bar{p}-\bar\tau^s)}{\sqrt{5 \pi}(32+23 \visrat)}\\
	&=\frac{4 \sqrt{5 \pi}  \left[-4 \Sr P_\ins^2+\left(-2+P_\out\right)^2\right]}{5(32+23\visrat)}\\
	&= \frac{9\sqrt{5 \pi}\exp\left(-\frac{4 \Rr t}{\delta_\mem(2+\Rr)}\right)}{5(32+23\visrat)(2+\Rr)^2}\\
	&\times \left[\left(\Rr+(2+\Rr) \exp\left(\frac{2 \Rr t}{\delta_\mem(2+\Rr)}\right)\right)^2-16 \Sr\right]\,,
\end{split}
\end{equation}
where $\bar p $ and $\bar \tau^s$ are the amplitudes of the electric pressure and the tangential  electric stress at the membrane for an electric field in the $z$-direction, see \refeq{pressure}--\refeq{tt} with $\alpha=0$, and $\beta=1$. Similar expressions for the electric pressure and tangential stress arising from an electric field in the $y$ direction are found in the appendix.

The term proportional to $\Ca^{-1}$  in \refeq{ev eq} describes the relaxation of the
 shape by bending stresses and the isotropic part of the membrane tension
\begin{equation}
\label{R2}
R_2=\frac{24(6+\Sigma_h)}{32+23 \visrat}\,.
\end{equation}
The membrane tension $\Sigma_h$ depends on the instantaneous vesicle shape and is determined self--consistently with deformation to keep the total area constant \cite{Vlahovska:2007},  see for details Appendix \ref{ap:solution}.  The leading order shape evolution equation becomes quadratic in the shape parameter $f$  in contrast to the corresponding results for  drops \cite{Taylor:1966}. This feature of non-equilibrium vesicle dynamics has been noted by several authors in relation to vesicle dynamics in shear flow \cite{Misbah:2006, Vlahovska:2007, Lebedev:2008}.

Inserting \refeq{R2} and  the expression for the tension $\Sigma_h$ \refeq{tens} in \refeq{ev eq} leads to
\begin{equation}
\label{f2m evolution}
\begin{split}
	\frac{\partial f_{20}}{\partial t}=C^\el_{20}(t)(1-2 \Delta^{-1}f^2_{20})\,,\\
	\frac{\partial f_{2n}}{\partial t}=-2C^\el_{20}(t) \Delta^{-1}f_{20}f_{2n}\,.
\end{split}
\end{equation}
The $f_{2n}$ modes are slaved to the $f_{20}$ shape mode,
which is forced to change by the electric field. An analytic solutions for $f_{20}$ can be found from the first equation in \refeq{f2m evolution},
\begin{align}\label{f20analytic}
	f_{20}(t) = \sqrt{\frac{\Delta}{2}}\tanh\left[ \sqrt{2\Delta}\left(D_{20}+J(t)\right) \right],
\end{align}
where $D_{20}$ is an integration constant determined from the initial conditions. Solving for the $f_{2\pm2}$  yields
\begin{align}\label{f22analytic}
	f_{2\pm2}(t) = D_{2\pm2} \text{sech}\left[ \sqrt{2\Delta}\left( \tilde{D}_{20}+\Delta^{-1} J(t) \right) \right],
\end{align}
where $D_{2\pm2}$ are determined from the intital conditions  and $\tilde{D}_{20}=D_{20}-J(1)(1-\Delta^{-1})$, and
\begin{align}
    J(t) = B_1\left\{t +\exp(-2\xi t)\left[B_2-2\delta_\mem\exp(\xi t)\right]\right\}\,.
\end{align}
where $B_1 = 9\sqrt{5\pi}/[5(32+23\eta)]$, $\xi = 2\Lambda/[\delta_\mem(2+\Lambda)]$, and $B_2 = (16S-\Lambda^2)/[2\xi(2+\Lambda)^2]$.
These solutions show that the maximum possible deformation of $f_{20}$ is found by letting $t\rightarrow\infty$,
\begin{equation}
\label{delta}
	f_{20}^{max}=\sqrt{\frac{\Delta}{2}}\,,
\end{equation}
which simply states that all excess area is transferred into the $f_{20}$ mode. A positive $f_{20}$ is characteristic of the prolate configuration, while when $f_{20}<0$, the vesicle is in the oblate configuration.
Equations \refeq{f20analytic} and \refeq{f22analytic} (or just \refeq{f2m evolution}) also show that the type of deformation can only change from oblate to prolate if $C^\el_{20}(t)$ changes sign. Setting $C^\el_{20}(t)=0$ and solving for the time $t$ we find that
\begin{align}\label{tob}
	t_{\text{ob}} = \frac{\delta_{\mem}(2+\Rr)}{2\Rr}\log\left[\frac{4\sqrt{\Sr}-\Rr}{2+\Rr}\right]
\end{align}
Since time can not be negative, the argument of the log function has to be greater than 1. Hence, this condition shows that a change in shape is possible if
 \begin{equation}
\label{Rc}
	\Rr<2\sqrt{\Sr}-1\,.
\end{equation}
If $\Sr=1$ , which is the typical case for vesicle experiments, we find that $\Rr<1$ in order to have oblate--prolate transition. This conclusion also follows by looking at the small time behavior of \refeq{f2m evolution} or \refeq{f20analytic}. One finds that the initial slope of $f_{20}$, for $\Sr=1$, will be positive when $\Rr>1$, and negative when $\Rr<1$. Equation \refeq{tob} also shows that increasing the membrane capacitance $\delta_{\mem}$ increases the time the vesicle spends in the oblate state. Fixing $\Sr$ and $\delta_{\mem}$ and decreasing $\Rr$ towards zero, also increases $t_{\text{ob}}$.

\begin{figure}
\begin{center}
\includegraphics[width=3.25in]{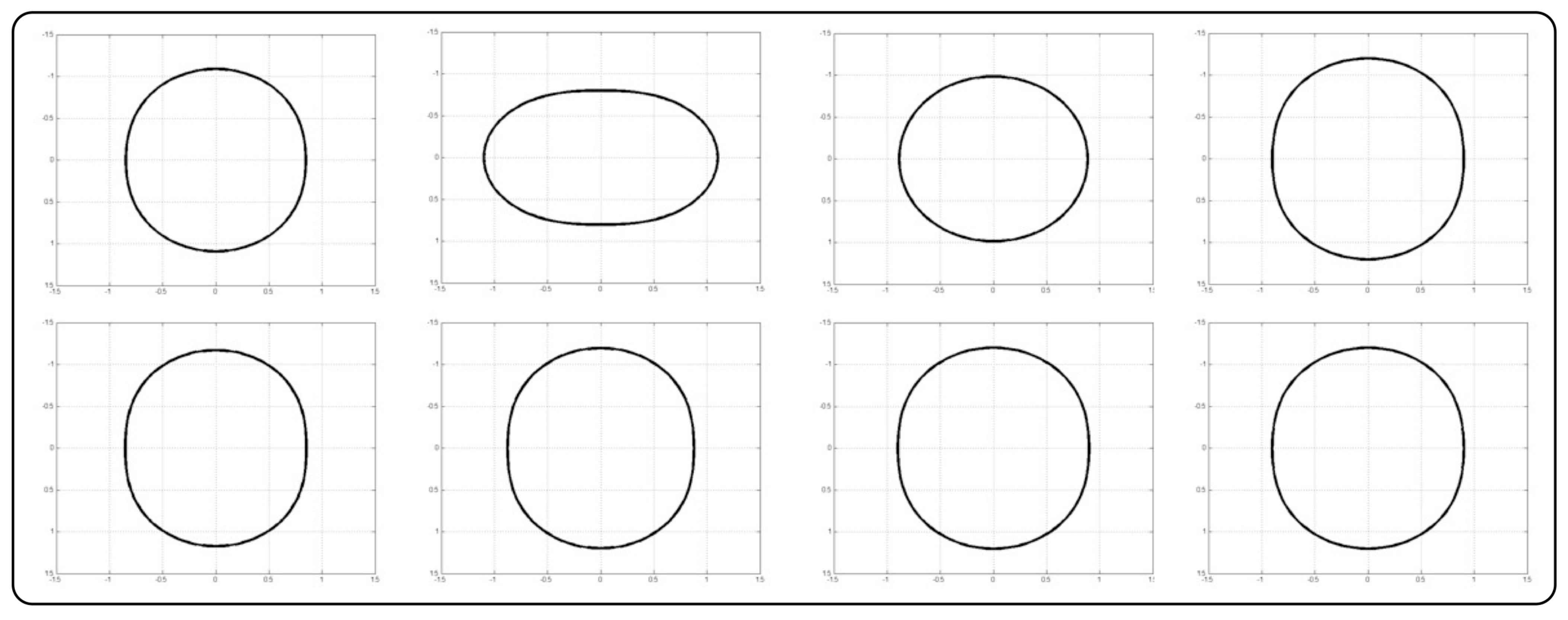}
\end{center}
\begin{picture}(0,0)(0,0)
\put(-95,70){$t=0$}
\put(-42,70){$t=2.7$}
\put(16,70){$t=8.5$}
\put(70,69){$t=15.5$}
\put(-95,25){$t=0$}
\put(-42,25){$t=0.25$}
\put(16,25){$t=1.0$}
\put(70,25){$t=10$}
\end{picture}
\caption{Contours of the vesicle shape in the $x-z$ plane. The top row is for a conductivity ratio of $\Rr=10$, and the bottom is for $\Rr=0.1$. Both solutions are with $\Pi=1$ and $\eta=1$. The initial conditions for the $f_{2\pm2} = f_{22}'\pm\im f_{22}''$ are $f_{22}'(0)=f_{22}''(0)=\sqrt{0.1\Delta}$. The $f_{20}$ mode is determined from Eq.\refeq{excessarea} with $\Delta=0.2$.}
\label{fig_shapes}
\end{figure}
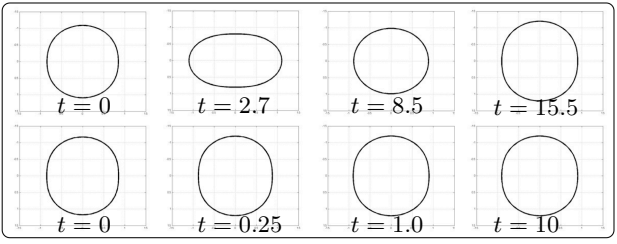

Figure \ref{fig_shapes} illustrates the time dependent shape dynamics obtained from \refeq{f2m evolution} upon a step-wise application of an uniform DC field. Initially, the excess area is distributed over all $f_{2n}$ modes. Over time, however,   the excess area is transferred into the $f_{20}$ mode. We observe that if $\Rr<1$ (and $\Sr=1$), the vesicle deforms initially into an oblate ellipsoid and then into a prolate ellipsoid.

\begin{figure}[h]
\centerline{\includegraphics[width=3in]{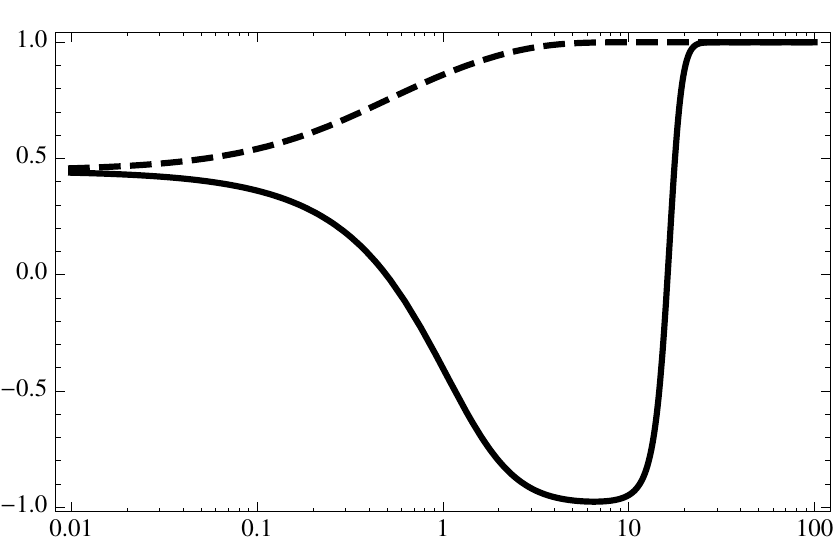}}
\vspace{0.25cm}
\centerline{\includegraphics[width=3in]{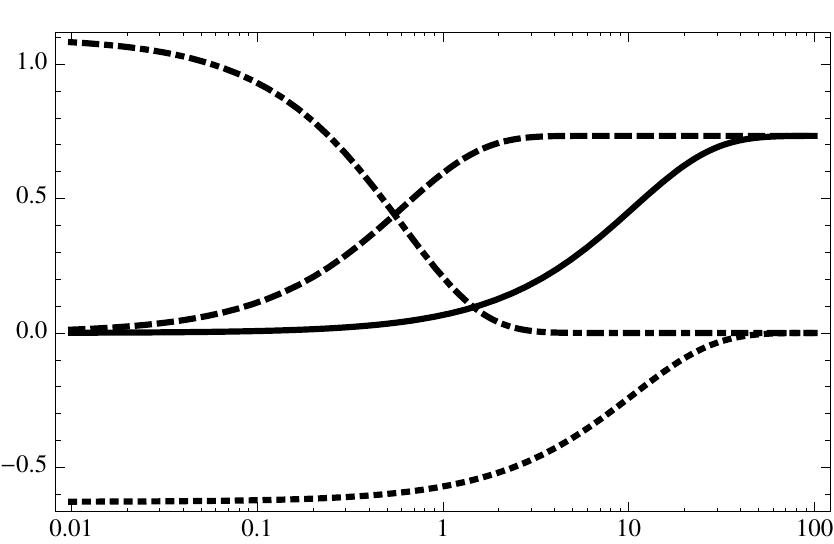}}
\vspace{0.25cm}
\centerline{\includegraphics[width=3in]{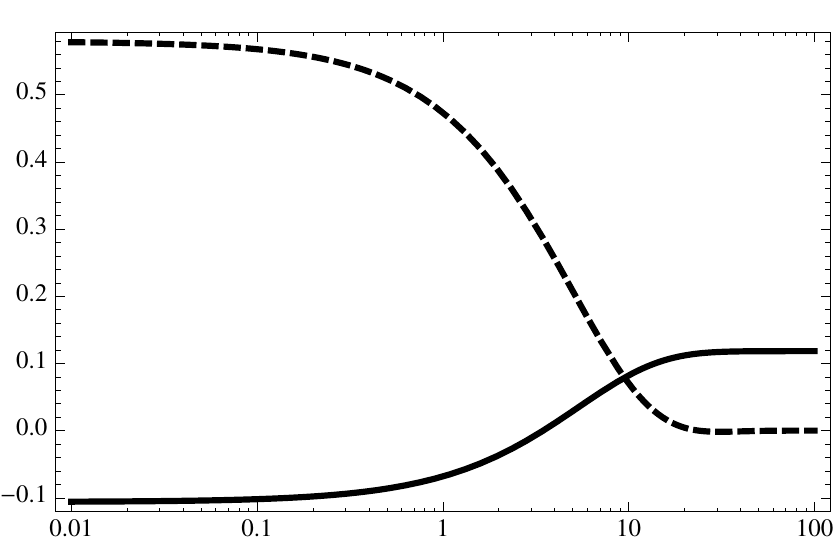}}
\begin{picture}(0,0)(0,0)
\put(-85,380){{{(a)}}}
\put(-85,250){{ {(b)}}}
\put(-85,120){{ {(c)}}}
\put(0,380){$\longleftarrow \Lambda=0.1$}
\put(-70,420){$ \Lambda=10\longrightarrow$}
\put(60,240){$\longleftarrow V_n(t)$}
\put(20,200){$Q(t)\longrightarrow$}
\put(30,100){$\longleftarrow \bar \tau^s $}
\put(-20,40){$ \bar p \longrightarrow$}
\put(-120,330){\rotatebox{90}{shape elongation $f_{20}/f_{20}^{max}$}}
\put(-120,170){\rotatebox{90}{transmembrane potential $\bar V$}}
\put(114,172){\rotatebox{90}{induced membrane charge $Q$}}
\put(114,35){\rotatebox{90}{Normal stress jump $\bar p$}}
\put(-120,25){\rotatebox{90}{Tangential stress jump $\bar \tau^s$}}
\put(-45,0){dimensionless time $t/t_{\cp}$}
\end{picture}
\caption{\footnotesize (a) Evolution of the ellipsoidal deformation $f_{20}/f_{20}^{max}$
of  a quasi-spherical vesicle upon application of a uniform DC electric field. The solid and dashed curves are with $\Lambda = 0.1$, and $10$, respectively.
(b) Evolution of the transmembrane potential (solid line ($\Lambda = 0.1$) and dashed line ($\Lambda = 10$)), computed from \refeq{VmpotDC}, and the effective charge (dotted line ($\Lambda=0.1$) and dotdashed line ($\Lambda=10$)) at the north pole. (c) electric pressure (solid line)  and shear stress (dashed line). Parameter values are $\Lambda=0.1$, $S=1$. Time is nondimensionalized by the capacitor charging time.}
\label{pulseshape}
\end{figure}
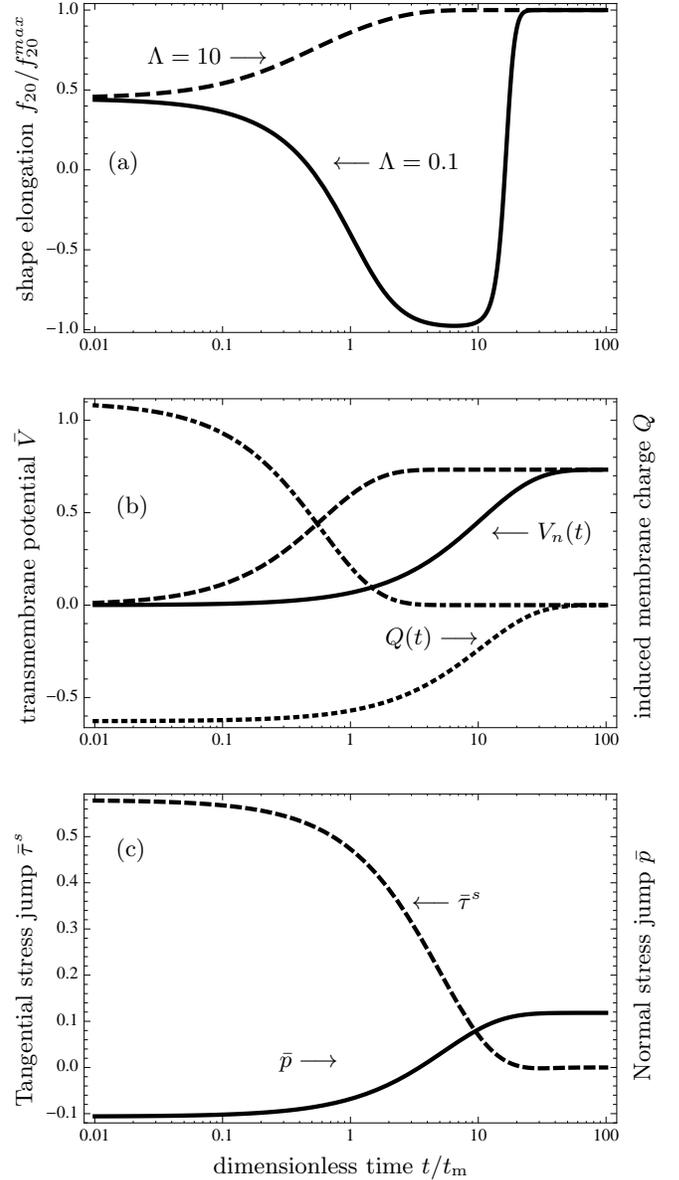

Figure \ref{pulseshape} illustrates the time-evolution of the $f_{20}$ shape mode, the induced charge and the transmembrane potential.  In the case $\Rr<1$, initially the induced effective surface charge is non-zero and the deformation is oblate-type. As time progresses and the membrane capacitor charges, the imbalance in the charge densities on the two membrane surfaces diminishes. Once the capacitor becomes fully charged, the effective surface charge vanishes, and the transmembrane potential reaches its steady state maximum value. The interior electric fields also vanishes and the vesicle assumes a prolate shape. These results indicate that the vesicle shape may change type during the application of a long DC pulse, which is still to be experimentally confirmed.

The stationary solution shown in \refeq{delta} (which defines a prolate vesicle) was obtained at by examining long time behavior of \refeq{f20analytic}. It can also be derived by assuming that the time dependent forcing from the electric field, $C_{20}^\el(t)$ has reached its steady state, which makes Equations \refeq{f2m evolution} autonomous. Introducing the constant quantity,
$C_{20}^{\el\infty} = C_{20}^\el(t=\infty)$
setting the left hand side of \refeq{f2m evolution} to zero, and solving for $f_{20}$ yields \refeq{delta}. However, a second stationary solution also exists
\begin{align}\label{delta_minus}
	f_{20}^{min} = -\sqrt{\frac{\Delta}{2}}\,.
\end{align}
This solution corresponds to an oblate spheroid. Linearizing the system \refeq{f2m evolution} about  the prolate solution \refeq{delta} and performing a stability analysis reveals real, negative eigenvalues ($\omega$),$\omega_1 = -C_{20}^{\el\infty}\sqrt{\Delta/2}$ and $\omega_2 = 2\omega_1$. This indicates that ,the prolate state is characterized by a stable node at long times. On the other hand, if the evolution equations \refeq{f2m evolution} were linearized about about the oblate state \refeq{delta_minus}, the eigenvalues are positive which this shows that at long times, the oblate state is an unstable solution.

Lastly, if one were to freeze time in $C_{20}^\el(t)$ at $t=0$, and not the long time state, the signs on the eigenvalues switch depending on the value of $\Rr$. In particular, if $\Rr<1$ and $C^{\el}_{20}(0)$ is negative, the equilibrium solution \refeq{delta_minus} is stable, and hence initially an attractor. On the other hand, if $\Rr>1$ and $C^{\el}_{20}(0)$ is positive, then \refeq{delta} is the attractor. These predictions are consistent with the dynamics shown in Figure \ref{pulseshape}.

\section{Vesicle dynamics in a combined electric field and shear flow }

The shape deformation modes $f_{2n}$ in a linear flow obey the evolution equation \cite{Vlahovska:2007}
\begin{equation}
\label{f22 evolution}
\frac{\partial f_{2n}}{\partial t}= \frac{\im n}{2}f_{2n}+C_{2n}-2\Delta^{-1}f_{2n}\sum_{n=-2}^{2}f_{2n}C_{2n}.
\end{equation}
In the presence of an electric field \cite{Schwalbe:thesis}
\begin{equation}
\label{eq:C-Mn}
	C_{2n}=C_{2n}^\shear+\Mn C_{2n}^\el\,.
\end{equation}
The forcing by the electric field is given by \refeq{Celforcing} and the contribution from the simple shear flow is \cite{Vlahovska:2007}
\begin{equation}
	C_{2n}^\shear=-\im n \frac{2 \sqrt{30 \pi}}{23\visrat+32}\,.
\end{equation}
For convenience when needed, the shape modes will be decomposed into their real and imaginary parts,
\begin{align}
	f_{jn} = f'_{jn} + \im f''_{jn}.
\end{align}
Instead of shape modes, the vesicle dynamics can be also conveniently described in terms of the orientation angle, $\psi$, and  $R$, which  measures the ellipticity of the vesicle contour in the $x-y$ plane \cite{Misbah:2006}
\begin{equation}
	f_{2\pm2}=R \exp(\mp2\im \psi)\,.
\end{equation}
The $f_{20}$ mode can be determined from the area constraint \refeq{excessarea}
\begin{equation}
\label{f20}
	f_{20}=\left[\frac{\Delta}{2}-2 f_{22} f_{2-2}\right]^{1/2}=\left[\frac{\Delta}{2}-2 R^2\right]^{1/2}\,.
\end{equation}
The evolution equations for the shape and orientation of a fluid membrane vesicle in a simple shear flow are \cite{Misbah:2006}
\begin{equation}
\label{psidot}
	\frac{\partial \psi}{\partial t}=-\frac{1}{2}-\frac{C''_{22}}{2 R(t)}\cos\left[2\psi(t)\right]-\frac{C_{22}'}{2 R	 (t)}\sin\left[2\psi(t)\right]\,,
\end{equation}
\begin{equation}
\label{Rdot}
\begin{split}
	\frac{\partial R}{\partial t}&=\left(1-4 \frac{R(t)^2}{\Delta}\right)\left\{C'_{22}\cos\left[2\psi(t)\right]- C''_{22}\sin\left[2\psi(t)\right]\right\}\\
&-2C_{20}R(t)\Delta^{-1}\left[\frac{\Delta}{2}-2 R^2\right]^{1/2}\,,
\end{split}
\end{equation}
where $C_{22}=C'_{22}+\im C''_{22}$. Note that in the absence of an electric field $(C'_{22}=0)$ and  $R$ constant, \refeq{psidot} reduces to the Keller-Skalak equation describing the dynamics of a tank--treading ellipsoid \cite{Keller-Skalak:1982}. According to this model, the TT state is characterized by a steady inclination angle; the transition from tank--treading to tumbling occurs when a steady--state solution for the inclination angle ceases to exist, i.e. $C''_{22}< R$. The electric field introduces a term phase-shifted; as a result \refeq{psidot} always has a TT solution.

\subsection{No electric field}
For the sake of completeness, here we summarize the results for vesicle dynamics in simple shear flow.
In the absence of electric field, $C_{20}=C'_{22}=0$, and $C''_{22}=-4\sqrt{30\pi}/(23 \eta+32)$ \cite{Misbah:2006}. In this case, a linear analysis of the set of coupled nonlinear equations, depending on the value of $\eta$,  results in a stable fixed point corresponding to the tank-treading state ($R^*=\sqrt{\Delta}/{2}\,, \cos(2\psi^*)=-\sqrt{\Delta}/{2C''_{22}}$) or a closed orbit centered at ($\psi^*=0\,, R^*=-C''_{22}$) describing the breathing mode.  Tumbling does not correspond to an equilibrium point. The TT fixed point loses stability at a critical viscosity ratio
\begin{equation}
\label{crit viscosity}
\visrat_c=-\frac{32}{23}+\frac{120}{23}\sqrt{\frac{2\pi}{15 \Delta}}\,.
\end{equation}

If there is no deformation along the vorticity direction, i.e.,  $f_{20}=0$ at all times, Eq. \refeq{f20} implies that  $R$ remains constant and equal to its maximum value $\sqrt{\Delta}/2$.
This situation resembles  the Keller-Skalak model \cite{Keller-Skalak:1982}: the vesicle shape is a fixed ellipsoid and the vesicle dynamics is described only by the variations of the angle $\psi$ (note, however, that unlike the Keller-Skalak solution, our velocity field is strictly area--incompressible). The nonlinear dynamics (either VB or TB) for $\eta>\eta_c$ will depend on the amplitude of the oscillation and the value of $\eta$. For example, in \cite{Vlahovska:2007}, results are presented which show a VB motion with large amplitude variation in $f_{20}$ for $\eta$ slightly greater than critical, while the dynamics become TB, with small amplitude variation of $f_{20}$, for values of $\eta$ much larger than critical. In the breathing mode, the vesicle undergoes periodic shape deformations along the vorticity direction and appears to tremble in the flow direction.

\subsection{Combined electric field and shear flow}
\subsubsection{Electric field along the velocity gradient direction}

\begin{figure}
\begin{center}
\hspace{.5cm}\includegraphics[width=3.2in]{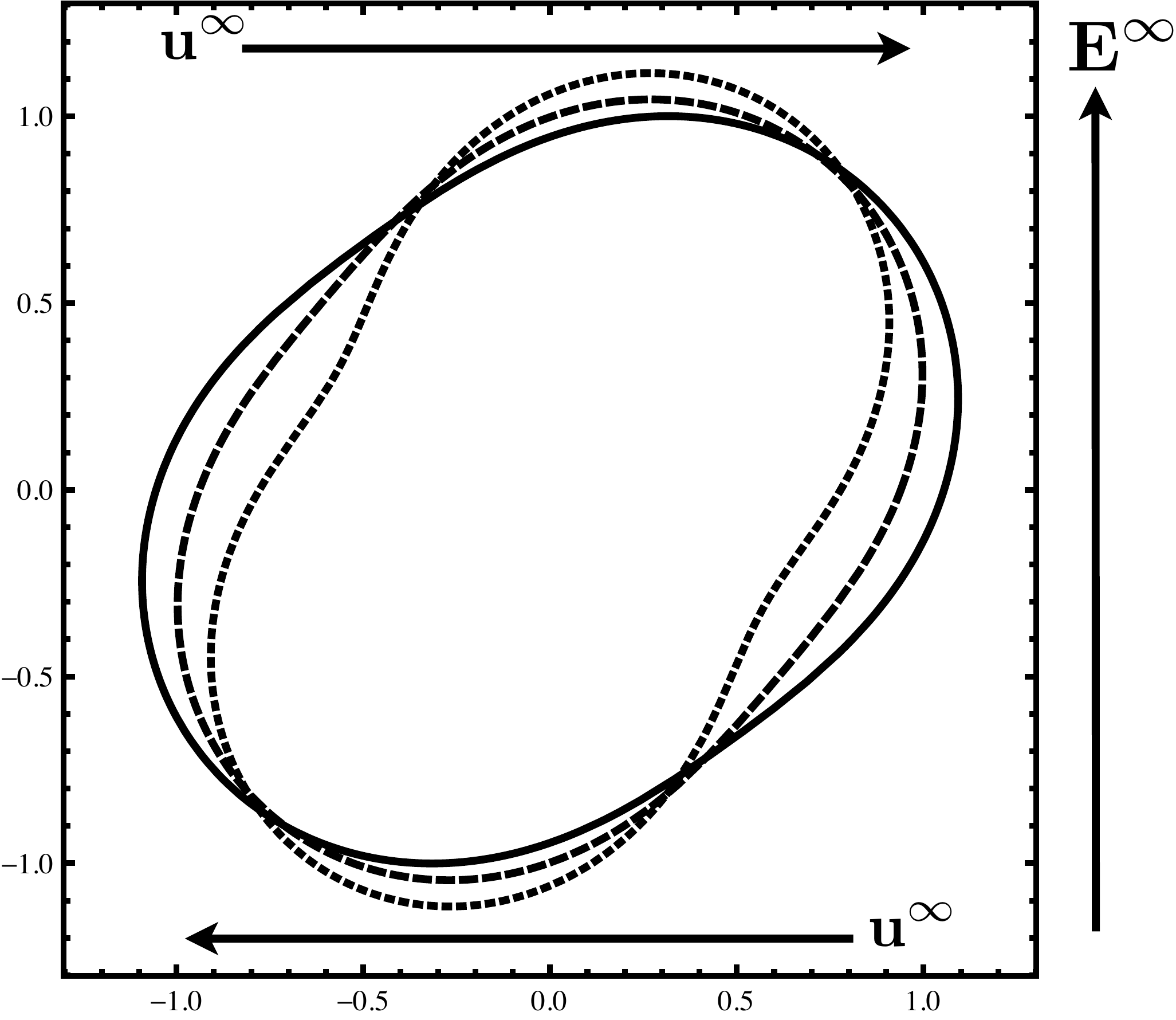}\\
\includegraphics[width=3.0in]{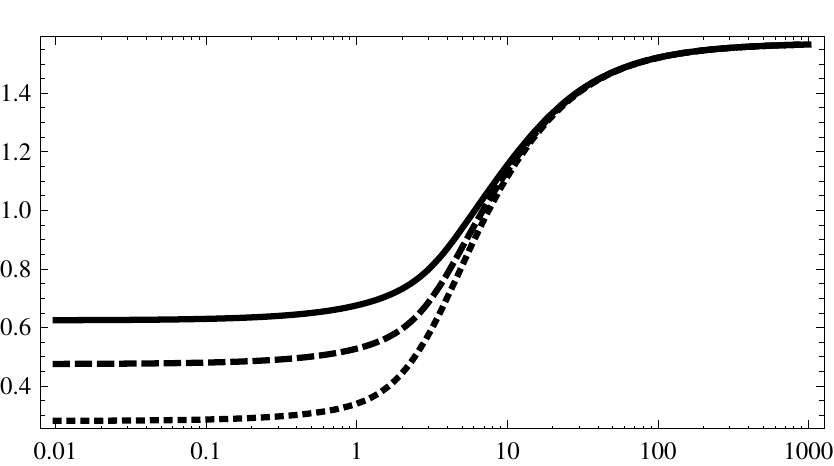}
\end{center}
\begin{picture}(0,0)(0,0)
\put(-85,300){(a)}
\put(-85,100){(b)}
\put(0,12){$\Mn$}
\put(-118,80){$\psi$}
\end{picture}
\caption{(a) Contours of the vesicle viewed in the  $x-y$ plane for $\eta=1$. The vesicle is stressed by the combined shear flow, and electric field in the $y$-direction ($\alpha=1$, $\beta=0$). The solid, dashed, and dotted contours are with $\Mn=0$, $\Mn=5$, and $\Mn=12$, respectively. The remaining parameters are $\Lambda=10$, $\Sr=1$, $\delta_{\mem}=10$, and $\Delta=0.2$. The initial conditions are the same as those in Figure \ref{fig_shapes}. (b) Inclination angle of vesicle at various $\Mn$. The solid, dashed and dotted curves are with $\eta = 1,3,$ and $5$, respectively.}
\label{tread-contours}
\end{figure}
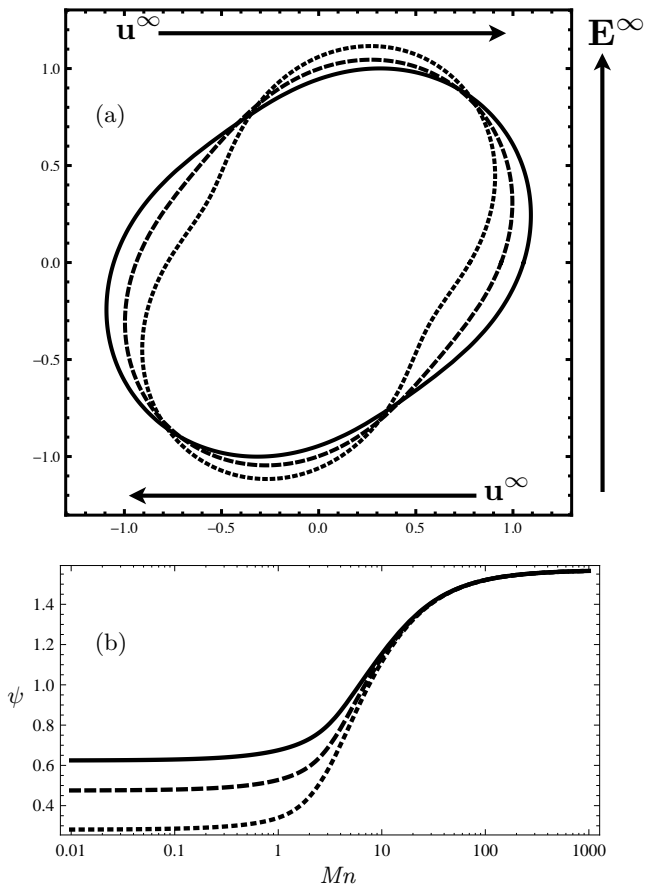

The presence of the electric field modifies the time dependent dynamics of the vesicle in shear flow. Consider first the case when the electric field is oriented in the $y$-direction. In the tank treading regime, the final state of the vesicle is influenced by the orientation, and strength of the electric field, as seen in Figure \ref{tread-contours}. The solution in this figure is found by integrating numerically \refeq{f22 evolution} and then using \refeq{radialcoordinate} and \refeq{expansions}.

The application of the electric field increases the inclination angle of the final state of the vesicle with respect to the flow direction. This effect can be seen in Figure \ref{tread-contours}.b. As $\Mn$ increases, $\psi$ increases towards $\pi/2$, the orientation of the applied electric field with respect to the $x$-axis.
This limiting value of $\psi$  can be obtained by letting $\Mn$ tend to infinity in  (\ref{eq:C-Mn}) and (\ref{psidot}). Note that as $\Mn$ increases, the value of $R$ tends to a finite value, but $C_{2n}$ increases in magnitude.  Hence (\ref{psidot}) implies that $\sin(2 \psi) = 0$.

\begin{figure}
\begin{center}
\includegraphics[width=3.0in]{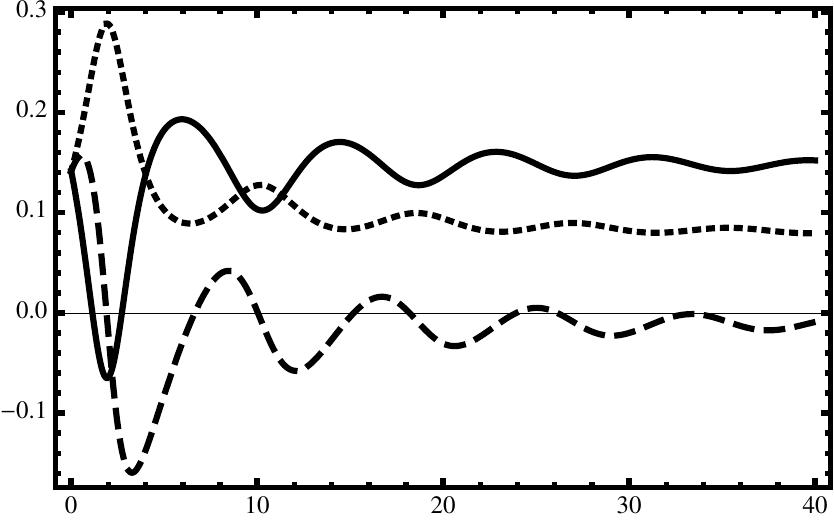}\\
\includegraphics[width=3.0in]{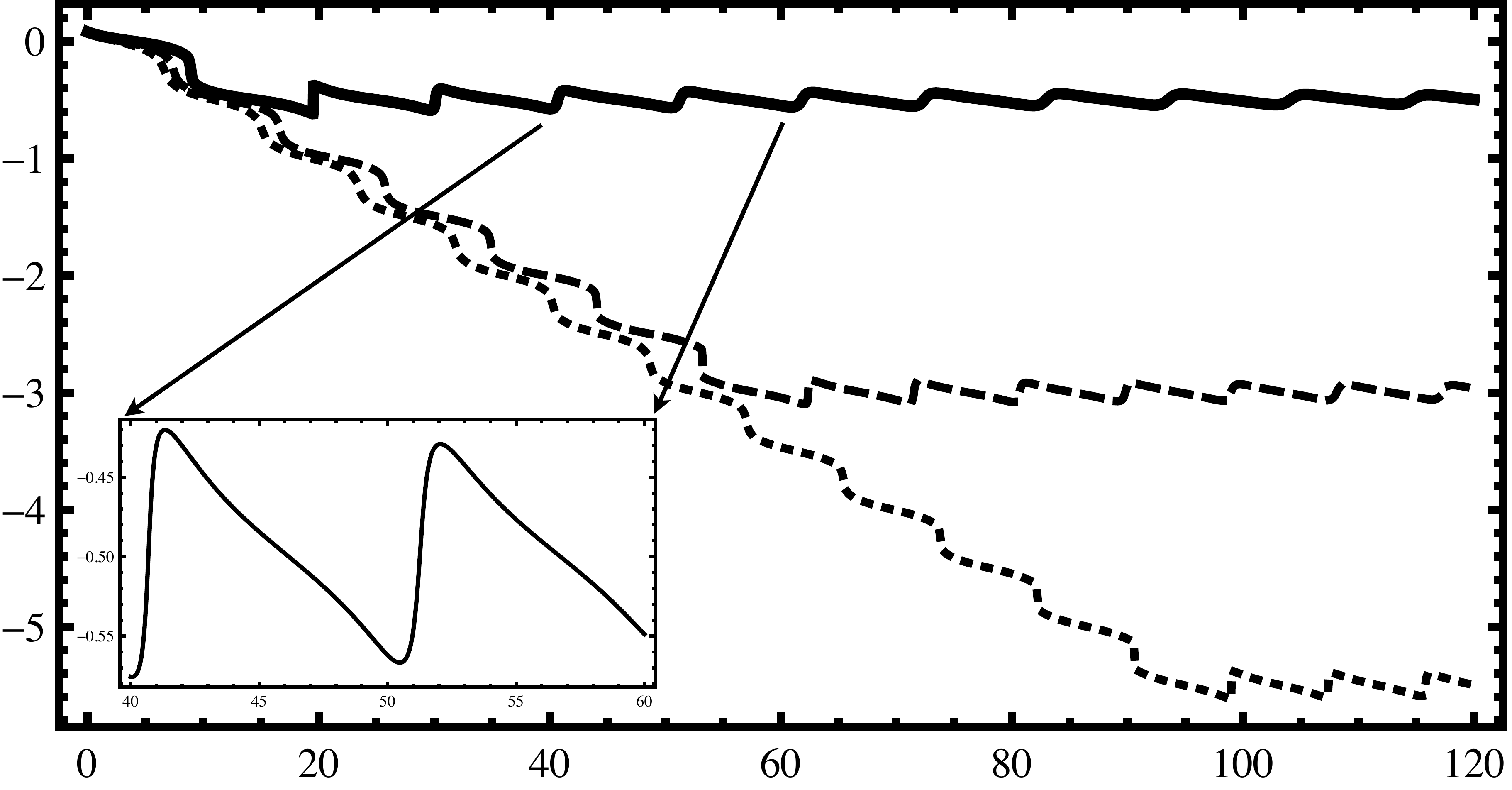}
\end{center}
\begin{picture}(0,0)(0,0)
\put(-45,10){dimensionless time $t/t_{\cp}$}
\put(-118,210){$f_{2n}$}
\put(-123,85){$\psi/2\pi$}
\put(70,240){(a)}
\put(70,105){(b)}
\end{picture}
\caption{(a) Time dependence of the $f_{2n}$ modes in the a damped tumbling state. The solid, dashed and dotted curves are $f_{22}''$, $f_{22}'$, and $f_{20}$, respectively with $\eta=10$. The remaining parameters are $\Rr=10$, $\delta_{\mem}=10$, $\Sr=1$, $\Mn=0.25$, and $\Delta=0.2$. (b) Time dependence of $\psi$. The solid, dashed and dotted curves are respectively with $\eta=8, 9$ and $10$. The reaming parameters are the same as those in part a.}
\label{damped_tumbling}
\end{figure}
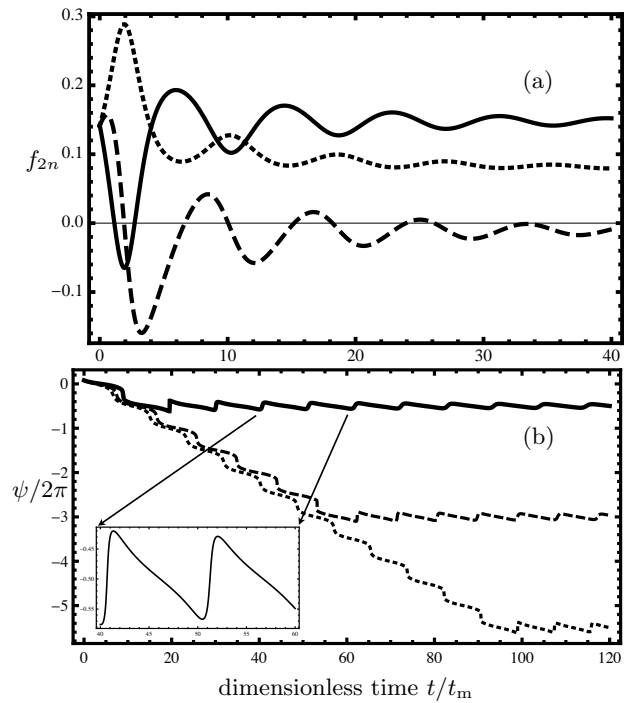

In the absence of the electric field, increasing the viscosity contrast $\eta$ brings the vesicle into the tumbling regime which is characterized by a periodic variation of the $f_{2n}$ modes. In the presence of an electric field, this periodic motion is damped. Figure \ref{damped_tumbling}.a illustrates  the vesicle tumbling  with decaying amplitude towards the tank treading configuration.
Increasing the viscosity of the vesicle, which enhances the viscous forces acting on the vesicle, only lengthens  the time required for the electric stresses to fully dampen the tumbling motion. This effect is seen in part b of Figure \ref{damped_tumbling} where $\psi$ is shown at various $\eta$. During the damped tumbling motion, the vesicle rotates counter-clockwise, and $\psi$ will increase negatively until the electric stresses have overcome tumbling motion, and the major axis can no longer make a complete rotation. The total number of rotations is therefore given by $\psi/2 \pi$.

A linear analysis similar to that of Part A of this section can be performed for $\Mn > 0$ in the long time limit where the time dependent coefficients are constant. Here the analysis is performed using the 3x3 system of equation \refeq{f22 evolution}. This choice was made in order to remain consistent with the discussion at the end of Section \ref{noflow}, where the system given by \refeq{f2m evolution} (dynamics described by shape modes) was used in the stability discussion. The steady state solutions can be identified, and a linear stability analysis can be done about these states which shows that only one of the solutions is stable.  This stable solution is a TT solution.  To illustrate this behavior, in Figure 7 we plot the eigenvalues (growth rates) associated with the linear stability of the stable TT solution for the special case of $\Mn = 1$, $\delta_m = 1$, $\Delta = 0.2$, $\Lambda =10$, and $S = 1$.  In Figure 7.a we plot the real part of the three eigenvalues while in Figure 7.b, we plot the imaginary parts.  Note that for $\eta$ less than some critical value $\eta_e$, there are three negative real eigenvalues, implying stability.  For $\eta > \eta_e$, two of the real eigenvalues become complex conjugates, with the real part remaining negative, still implying stability.  These numerical results show that the same steady state solution is stable for all values of $\eta$, but how this solution is approached differs as $\eta$ varies.  In particular, for small $\eta$ we see a stable node while if $\eta > \eta_e$, there is a damped oscillation into the tank treading solution.  It is interesting to note that the computed value of $\eta_e  \approx 4$ is less than the critical $\eta$ for the $\Mn = 0$ case given in equation (\ref{crit viscosity}), $\eta_c = 6.15$, implying the $\eta_e$ depends on the other physical parameters in addition to $\Delta$. Finally, note that this linear stability analysis yields results consistent with the numerical solution of the system as presented in Figures 5 and 6.

\begin{figure}
\begin{center}
\includegraphics[width=3.0in]{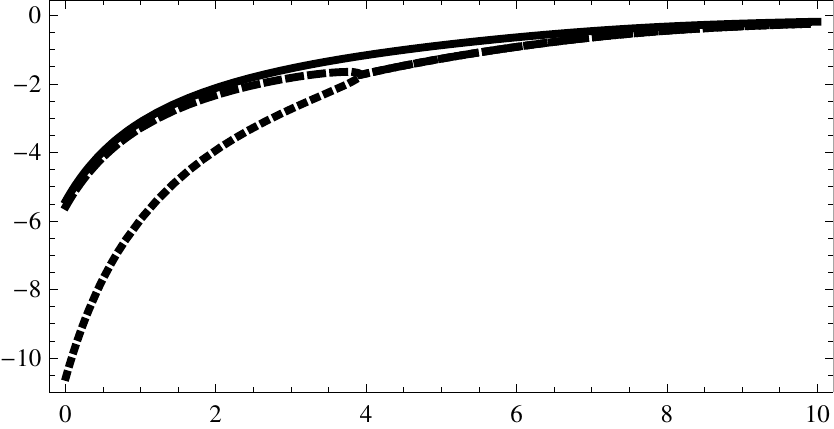}\\
\includegraphics[width=3.0in]{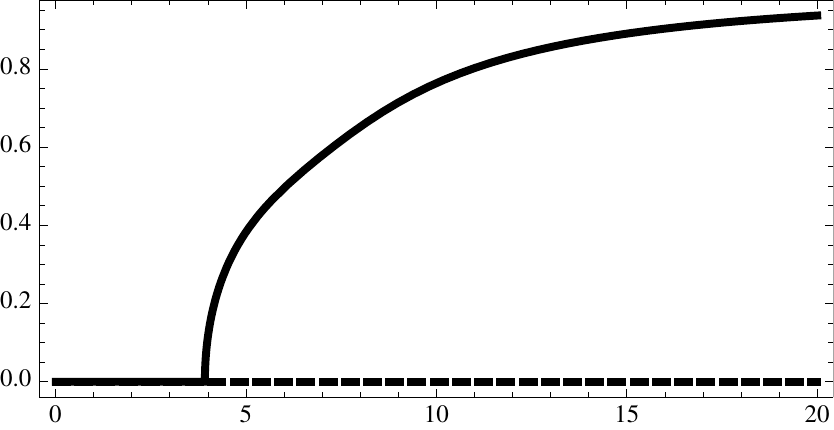}
\end{center}
\begin{picture}(0,0)(0,0)
\put(0,12){$\eta$}
\put(-120,165){\rotatebox{90}{Re[Eigenvalues]}}
\put(-120,50){\rotatebox{90}{Im[Eigenvalues]}}
\put(70,220){(a)}
\put(70,105){(b)}
\end{picture}
\caption{The magnitude of the real (a), and imaginary (b) components of the eigenvalues of the system \refeq{f22 evolution} linearized about its steady state, as a function of $\eta$. In (b), the solid, and dashed curves represent the magnitude of the two unique imaginary components. The remaining parameters are $\Rr =10$, $S=1$, $\delta_m=1$, and $\Mn=1$.}
\label{Eigs}
\end{figure}
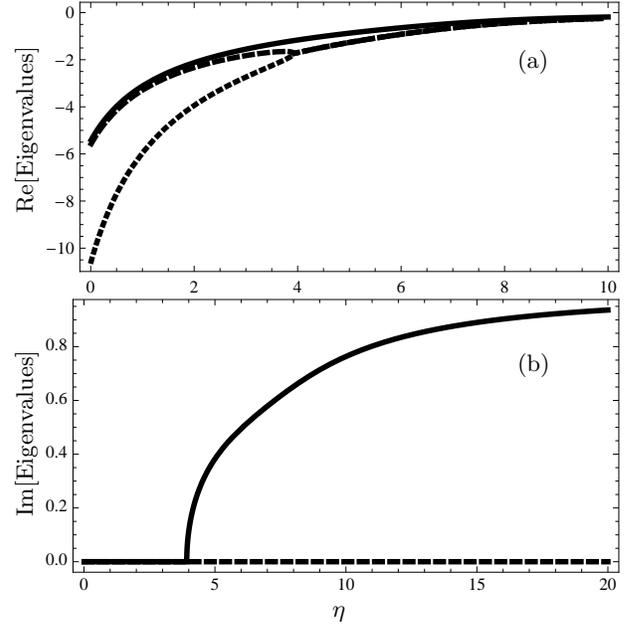

Steady states for $R$ and $\psi$ can be found from \refeq{psidot} and \refeq{Rdot}. Upon using \refeq{psidot} to obtain an expression for $R$ in terms of $\psi$ and inserting it into \refeq{Rdot}, one obtains a fourth order polynomial equation for $R$,
\begin{align}\label{R:poly}
	R^4-\left(|C_{22}|^2 + \textstyle{\frac{1}{4}}\Delta + \textstyle{\frac{1}{2}}C_{20}^2\right)R^2+\textstyle{\frac{1}{4}}\Delta |C_{22}|^2=0.
\end{align}
where $|C_{22}|^2 = (C_{22}'')^2 + (C_{22}')^2$. Eq. \refeq{R:poly} can be solved to give $R^2$,
\begin{align}\begin{split}\label{rss}
    R^2 =& \textstyle{\frac{1}{2}}|C_{22}|^2 + \textstyle{\frac{1}{8}}\Delta + \textstyle{\frac{1}{4}}C_{20}^2\\
    &\pm\textstyle{\frac{1}{4}}\left[ (2|C_{22}|^2+\textstyle{\frac{1}{2}}\Delta +C_{20}^2)^2-4\Delta |C_{22}|^2 \right]^{1/2}.
\end{split}\end{align}
From this, steady states of $\psi$ can be found from \refeq{psidot}.

\subsubsection{Electric field along the vorticity direction}
Orientation of the electric field along other coordinate axes results in similar time dependent dynamics of the shape modes seen in Figure \ref{damped_tumbling}, i.e., the vesicle undergoes a damped tumbling motion. We should note that when the electric field is directed along the $z$ coordinate axis ($\alpha=0$, $\beta=1$), see \refeq{Einf}, the system reduces to a more compact form. Steady states can be analytically obtained from \refeq{f22 evolution} by solving
\begin{align}
	0&= -\Delta f''_{22} + 2f'_{22}(2C_{22}^{\shear} f''_{22}-C_{20}^\el f_{20})\label{f20s}\\
	0&= \Delta(f'_{22} + C_{22}^{\shear}) + 2f''_{22}(2C_{22}^{\shear} f''_{22}-C_{20}^\el f_{20})\label{f22sr}\\
	0 &= 4 C_{22}^{\shear} f_{20} f_{22}'' + C_{20}^\el (\Delta - 2f_{20}^2)\label{f22si}
\end{align}
 for $f'_{22}$, $f''_{22}$, and $f_{20}$. Note that one must also assume the time dependent coefficients have reached their steady state. Details concerning the derivation of the steady states are left to the appendix; in the end, a fourth order polynomial for $f_{20}$ is found,
\begin{align}\label{f20poly}
	2f_{20}^4 + [2(C_{20}^\el)^2 + 4C_{22}^\shear-\Delta]f_{20}^2-\Delta (C_{20}^\el)^2=0.
\end{align}
The four steady states of $f_{20}$ are
\begin{align}\label{f201}
	f_{20} = \pm\frac{1}{2}\left[ -2(C_{20}^\el)^2-\gamma-\zeta \right]^{1/2}
\end{align}
and
\begin{align}\label{f202}
	f_{20} = \pm\frac{1}{2}\left[ -2(C_{20}^\el)^2+\gamma+\zeta \right]^{1/2},
\end{align}
where $\gamma = 4(C_{22}^\shear)^2 - \Delta$, and
\begin{align}
	\zeta = \left[ 8\Delta(C_{20}^\el)^2 + (2(C_{20}^\el)^2+\gamma)^2 \right]^{1/2}.
\end{align}
The two solutions given by \refeq{f201} are imaginary for physical values of the parameters, and therefore are not valid. The two solutions given by \refeq{f202} are real, and therefore valid solutions of the system. We linearize the evolution equations \refeq{f22 evolution} about the solutions given by \refeq{f202}, and with the corresponding $f'_{22}$ and $f''_{22}$ (found from \refeq{ref} and \refeq{imf} respectively). The eigenvalues this system are complex with either a positive or negative real part, depending on the sign taken for $f_{20}$ from \refeq{f202}; adopting the positive sign for $f_{20}$ yields a stable system.

\section{Conclusions}

In this study we considered the effects of a steady uniform electric field on the dynamics of a vesicle in a simple shear flow. We have adapted a model which accounts for the fluidity and incompressibility of the interface in addition to bending resistance. The interface was treated as a capacitor and thus the boundary conditions at the membrane have intrinsic time-dependence. In the limit of a nearly spherical vesicle and weak electric field, we derived a system of coupled nonlinear ordinary differential equations with time dependent coefficients which describe the evolution of the vesicle shape.

The solution of the evolution equations  shows that in the absence of an applied shear flow, the vesicle  either remains a prolate ellipsoid for all time, or temporarily enters a oblate state before becoming prolate.  Under the combined action of shear flow and electric field, at steady state the vesicle is a tank--treading prolate ellipsoid, which can be reached either monotonically or via damped tumbling. Our theoretical results are consistent with available experimental data \cite{Riske-Dimova:2006} albeit some of the theoretical predictions such as the oblate--prolate transition in absence of applied flow and the suppressed tumbling under shear when an electric field is present remain to be experimentally tested.

\section*{Acknowledgments}
JTS and MJM acknowledge financial support by NSF RTG grant DMS-0636574 and NSF grant DMS-0616468. PMV acknowledges partial financial support by NSF grant CBET-0846247.

\appendix

\section{Spherical harmonics}
\label{ap:harmonics}

The normalized scalar spherical harmonics are defined as,
\be
\label{normalized spherical harmonics}
	Y_{jn}(\theta,\phi) = \left[ \frac{2j+1}{4\pi} \frac{(j-n!)}{(j+n)!}\right]^\half(-1)^nP_j^n(\cos \theta)e^{\im n\phi},
\ee
where  $P_j^n(\cos\theta)$ are the associated Legendre polynomials.  For example
\begin{equation}
Y_{10}=\sqrt{\frac{3}{4 \pi}}\cos\theta \,.
\end{equation}
The vector spherical harmonics relevant to our study are defined as \cite[]{Blawzdziewicz-Vlahovska-Loewenberg:2000}
\bea\begin{split}
\label{vector harmonics}
	\mbf{y}_{jn0} &= \left[j(j+1)\right]^{-1/2}r\nabla_{\Omega}Y_{jn},\\
	\mbf{y}_{jn1} &= -\im\hat{\mbf{r}}\times\mbf{y}_{jn0},\\
	\mbf{y}_{jn2} &= \hat{\mbf{r}} Y_{jn}.
\end{split}\eea
For example
\begin{equation}
\label{vh2}
\bS_{200}=-\textstyle\sqrt{\frac{15}{32\pi}} \sin(2 \theta)\bs{e}_\theta,\quad
\bS_{202}=\textstyle \frac{1}{8}\sqrt{\frac{5}{\pi}}[1+3\cos(2\theta)]\rhat\,.
\end{equation}

\begin{equation}
\bS_{222}+\bS_{2-22}=\sqrt{\frac{15}{8\pi}}  \left(\cos 2 \phi  \sin ^2\theta\right)\rhat
\end{equation}

\begin{equation}
\bS_{220}+\bS_{2-20}=\sqrt{\frac{5}{4\pi}}\left[\half( \cos 2\phi\sin2\theta){\bs{e}_\theta}-(\sin2 \phi\sin\theta){\bs{e}_\phi}\right]
\end{equation}

Calculations of the electric tractions involve recoupling of products of vector and scalar spherical harmonics. A detailed presentation of general recoupling formulas is beyond the scope of this paper and can be found in \cite{Vlahovska:2005}. Here we  list the formulas needed to complete the calculation in this work.
\begin{subequations}
\begin{align}\begin{split}
	{Y}_{1\pm1}Y_{1\pm1} &= \sqrt{\frac{3}{10\pi}}{Y}_{2\pm20},\\
	{Y}_{1-1}Y_{11} &= -\frac{1}{2\sqrt{\pi}}Y_{00}+\frac{1}{2\sqrt{5\pi}}{Y}_{20}.
\end{split}\end{align}
\begin{align}\begin{split}
	\sqrt{2}Y_{1\pm1}\mbf{y}_{1\pm10}&=\frac{3}{2\sqrt{10\pi}}\bS_{2\pm2},\\
	\sqrt{2}Y_{1\pm1}\mbf{y}_{1\mp10}&=\mp\frac{1}{2}\sqrt{\frac{3}{2\pi}}\bS_{101} +  \frac{1}{4}\sqrt{\frac{3}{5\pi}}\bS_{200}\label{crosses}.
\end{split}\end{align}
\begin{align}\begin{split}
	2\mbf{y}_{1\pm10}\cdot\mbf{y}_{1\pm10} &= -\sqrt{\frac{3}{10\pi}}{Y}_{2\pm2},\\
	2\mbf{y}_{1-10}\cdot\mbf{y}_{110} &=-\frac{1}{\sqrt{\pi}}{Y}_{00} + \frac{1}{\sqrt{5\pi}}{Y}_{20}
\end{split}\end{align}
\end{subequations}

\section{Inhomogeneous forcing from $y$ electric field}
Here we list the time dependent forcing terms, $C_{2n}^\el$ resulting from an electric field in the $y$ direction
\begin{align}\begin{split}
	C_{22}^\el &= \sqrt{\frac{6\pi}{5}}\frac{\exp\left(-\frac{4\Rr t}{\delta_\mem(2+\Rr)}\right)}{(32+23\eta)(2+\Rr)^2}\\
	&\times\Bigg[9(\Rr^2-5+4\Sr)+(2+\Rr)\exp\left(\frac{2\Rr t}{\delta_\mem(2+\Rr)}\right)\\
	&\times\left(6-9\Rr-2(2+\Rr)\exp\left(\frac{2\Rr t}{\delta_\mem(2+\Rr)}\right)\right)\Bigg],
\end{split}\end{align}
and
\begin{align}\begin{split}
	&C_{20}^\el = \frac{9\sqrt{\pi}\exp\left(-\frac{4\Rr t}{\delta_\mem(2+\Rr)}\right)}{2\sqrt{5}(32+23\eta)(2+\Rr)^2}\\
	&\times\Bigg[ 16\Sr-\left(\Rr+(2+\Rr)\exp\left(\frac{2\Rr t}{\delta_\mem(2+\Rr)}\right)\right)^2 \Bigg].
\end{split}\end{align}

\section{Solution of for the electric field driven flow}
\label{ap:solution}

Here we outline the solution for the velocity field resulting from electric tractions in the case of a sphere placed in an uniform electric field. More details can be found in Refs.~\cite{Seifert:1999, Vlahovska:2007, Danker:2007b, Lebedev:2008, Schwalbe:2010}. The formalism was originally developed to study droplets in flow \cite{Vlahovska:2003, Vlahovska:2005, Vlahovska:2009a}.

Velocity fields are described using basis sets of fundamental solutions of the Stokes equations appropriate for spherical geometry \cite{Cichocki-Felderhof-Schmitz:1988}, $\bu^\pm_{jmq}$, defined in Section~\ref{Ap:velocity basis}:
\begin{equation}
\begin{split}
\label{velocity fields}
\bv^{\out}(\br)=\sum_{jmq}c_{jmq}\bu^{-}_{jmq}(\br)\,,\\
 \bv^{\ins}\brac{\br}=\sum_{jmq}c_{jmq}\bu^{+}_{jmq}(\br)\,.
 \end{split}
\end{equation}
\begin{equation}
\sum_{jmq}\equiv\sum_{j=2}^\infty\sum_{m=-j}^j\sum_{q=0}^{2}
\end{equation}
\begin{align}\label{Sigmaexpansions}
	\Sigma(\theta,\phi,t) &= \Sigma_h+ \sum_{j\geq2}\sum_{m=-j}^j\Sigma_{jm}(t) Y_{jm}\,,
\end{align}
where $\Sigma_h$ is the isotropic part of the tension used to enforce a global constraint on the area.
The
local area conservation implies that the velocity field at the
interface is solenoidal \cite{Seifert:1999}
\begin{equation}
\label{membrane incompressibility}
\bnabla_s \cdot \bv=0\,.
\end{equation}
Therefore the amplitudes of the velocity field \refeq{velocity fields} are related
\begin{equation}
\label{solen}
c_{jm0}=\frac{2}{ \sqrt{j(j+1)}}c_{jm2}\,.
\end{equation}
The component of velocity that is normal to the interface, $c_{jm2}$, is determined using the stress balance, which  in terms of spherical harmonics  reads
\begin{equation}
\label{stress bal2}
 \delta_{j2}\delta_{m0}\tau^{\el}_{jmq}+\tau^{\hd,\out}_{jmq}-\visrat \tau^{\hd,\ins}_{jmq}=\Ca^{-1}\tau^\mem_{jmq} \,.
\end{equation}
Tangential stresses correspond to the $q=0$ component, and the normal stresses - to  $q=2$. $\delta_{ij}$ is the Kronecker delta function. The hydrodynamic tractions are given by  \refeq{HD trac:1}--\refeq{HD trac:2}. The electrical tractions  are given by (see Section \ref{ves:stress})
\begin{equation}
{{\tau}}^\el=8\sqrt{\frac{\pi}{5}}p^\el\bS_{202}(\theta,\phi)  -2\sqrt{\frac{2\pi}{15}}\tau_s^\el \bS_{200}(\theta,\phi) \,.
\end{equation}
 The membrane tractions are \cite{Vlahovska:2007,Seifert:1999}
\begin{equation}
\tau^\mem_{jmq}=\tau^\kappa_{jmq}+\tau^\Sigma_{jmq}\,.
\end{equation}
The bending  contribution to the membrane traction is
\begin{equation}
\tau^\kappa_{jm2}=\textstyle j(j+1)\left(j-1\right)\left(j+2\right)f_{jm} \,, \quad
\tau^\kappa_{jm0}=0\,,
\end{equation}
the stresses due to membrane tension are
\begin{equation}
\begin{split}
\tau^\Sigma_{jm2}=&\textstyle 2\Sigma_{jm}+\Sigma_h\left(j-1\right)\left(j+2\right)f_{jm}\,, \\
\tau^\Sigma_{jm0}=&-\sqrt{j(j+1)}\Sigma_{jm}\,.
\end{split}
\end{equation}
The non-uniform part of the membrane tension, $\Sigma_{jm}$, is determined from the tangential component  of the stress balance \refeq{stress bal2}, $q=0$,
\begin{equation}
\Sigma_{jm}=\Ca\left[\frac{\tau^\el_{jm0}}{\sqrt{j(j+1)}}+ c_{jm2}\frac{2+j+(j-1)\visrat}{j(j+1)}\right]\,.
\end{equation}
It is then substituted into the normal component of the stress balance \refeq{stress bal2}, $ q=2$,  to obtain the normal velocity $c_{jm2}$
\begin{equation}
\label{ca}
c_{jm2}=C_{jm}+\Ca^{-1}(\Gamma_1+\Sigma_h\Gamma_2)f_{jm}\,,
\end{equation}
where
\begin{equation}
\label{Cjm}
\begin{array}{ll}
C_{jm}=-\frac{\sqrt{j(j+1)}}{d(\visrat,j)}\left[2\tau^\el_{jm0}+\sqrt{j(j+1)}\tau^\el_{jm2}\right]\,,
\end{array}
\end{equation}
\begin{equation}
\label{Gamma1}
\Gamma_1=-(j+2)(j-1)[j(j+1)]^2{d(\visrat,j)}^{-1}\,,
\end{equation}
\begin{equation}
\label{Gamma2}
\Gamma_2=-(j+2)(j-1)j(j+1){d(\visrat,j)}^{-1}\,,
\end{equation}
and
\begin{equation}
d(\visrat, j)=(4+3j^2+2j^3)+(-5+3j^2+2j^3)\visrat\,.
\end{equation}

Finally, the motion of the interface is determined from the kinematic condition \refeq{interfaceevolution}
\begin{equation}
\label{interface evolution:2}
\frac{\partial f_{jm}}{\partial t}= c_{jm2}+\frac{\im m}{2} f_{jm}\quad {\mbox{at}}\,\, r=1\,.
\end{equation}
Substituting $c_{jm2}$ in \refeq{interface evolution:2} yields the evolution equation  for the  shape parameters \refeq{ev eq}.
\begin{equation}
\label{tension}
\Sigma_h=-\frac{\sum_{jm} a(j) \left[C_{jm}f^*_{jm}+ \Ca^{-1} \Gamma_1 f_{jm}f^*_{jm}\right]}{\Ca^{-1}_\ehd\sum_{jm} a(j) \Gamma_2 f_{jm}f^*_{jm}}\,.
\end{equation}
The normal velocity \refeq{ca} and the shape evolution \refeq{interface evolution:2} include the yet unknown  isotropic membrane tension.
It is expressed  in terms of the shape modes and other known parameters in the problem using the area constraint \cite{Vlahovska:2007}

The complicated dependence of the tension on the shape modes makes the shape evolution equations  nonlinear.

In order to clarify the physical significance  of the isotropic tension, let us consider  the particular case when only  the ellipsoidal deformation modes, $j=2$, are present. \refeq{tension} simplifies to
\begin{equation}\begin{split}
\label{tens}
\Sigma_h(t)=&-6+\Ca \frac{32+23\visrat }{12}\big[ C_{20} f_{20}(t)\\
&+C_{22} f_{22}(t)+C_{2-2} f_{2-2}(t)\big]
\end{split}\end{equation}
where we have emphasized that the time dependent shape modes give rise to time-dependent membrane tension. We see that the tension varies with deformation.

In absence of applied shear, and electric field along the $z$-axis, once all  excess area is transferred to the $f_{20}$ mode, the tension increases with the field strength $\Ca$ as
\begin{equation}
\label{tens2}
\Sigma_h\approx \Ca C_{20}\frac{(32+23\visrat)\sqrt{2}}{12}\Delta^{-1/2}
\end{equation}
Similar behavior is observed with vesicles in shear flow \cite{Seifert:1999}.

\section{Steady state analysis}
In this section details concerning the derivation of the polynomial in \refeq{f20poly} will be shown. From \refeq{f20s}, the steady state of $f_{22}''$ is found to be
\begin{align}\label{imf}
	f''_{22} = \frac{C_{20}^\el(2f_{20}^2-\Delta)}{4C_{22}^\shear f_{20}}.
\end{align}
Additionally, using \refeq{f22sr}, and expression for $f_{22}'$ can be found,
\begin{align}\label{ref}
	f'_{22} = \frac{\Delta f''_{22}}{2C_{22}^\el f_{20} - 4C_{22}^\shear f''_{22}}.
\end{align}
Inserting \refeq{imf} into \refeq{ref}, and using \refeq{f22si} yields \refeq{f20poly}.

\section{Fundamental set of velocity fields}
\label{Ap:velocity basis}
Following the definitions given in Blawzdziewicz {\it et al.}\cite{Blawzdziewicz-Vlahovska-Loewenberg:2000}, we list the expressions for the functions ${\bu}^\pm_{jmq}\left(r, \theta,\varphi\right)$. The velocity field outside the vesicle is described by
\begin{equation}
\label{-vel 0}
\begin{array}{ll}
\bu^-_{jm0}={\textstyle\frac{1}{2}}r^{-j}\left(2-j+j r^{-2}\right)\bS_{jm0}+\\
\hspace{1cm}{\textstyle\frac{1}{2}}r^{-j}\left[j\left(j+1\right)\right]^{1/2}\left(1-r^{-2}\right) \bS_{jm2}\,,
\end{array}
\end{equation}
\begin{equation}
\label{-vel 2}
\begin{array}{ll}
\bu^-_{jm2}={\textstyle\frac{1}{2}}r^{-j}\left(2-j\right) \left(\frac{j}{1+j}\right)^{1/2}\left(1-r^{-2}\right)\bS_{jm0}+\\ \hspace{1cm}{\textstyle\frac{1}{2}}r^{-j}\left(j+(2-j)r^{-2}\right)\bS_{jm2}\,.
\end{array}
\end{equation}
The velocity field inside the vesicle is described by
\begin{equation}
\label{+vel 0}
\begin{array}{ll}
\bu^+_{jm0}= {\textstyle\frac{1}{2}}r^{j-1}\left(-(j+1)+(j +3)r^2\right)\bS_{jm0}-\\
\hspace{1cm}{\textstyle\frac{1}{2}}r^{j-1}\left[j\left(j+1\right)\right]^{1/2}\left(1-r^2\right)\bS_{jm2}\,,
\end{array}
\end{equation}
\begin{equation}
\label{+vel 2}
\begin{array}{ll}
\bu^+_{jm2}={\textstyle\frac{1}{2}}r^{j-1}\left(3+j\right)\left(\frac{j+1}{j}\right)^{1/2}\left(1-r^2\right)\bS_{jm0}+\\ \hspace{1cm}{\textstyle\frac{1}{2}}r^{j-1}\left(j +3-(j+1)r^2  \right)\bS_{jm2}\,.
\end{array}
\end{equation}
On a sphere $r=1$ these velocity fields reduce to the vector spherical harmonics defined by \refeq{vector harmonics}
\begin{equation}
\bu^{\pm}_{jmq}=\bS_{jmq}\,.
\end{equation}
Hence, $\bu^{\pm}_{jm0}$ is tangential,  and $\bu^{\pm}_{jm2}$ is normal to a sphere. In addition,  $\bu^{\pm}_{jm0}$ defines an irrotational velocity field.

The hydrodynamic tractions associated with the velocity fields \refeq{velocity fields} are \cite{Vlahovska:2007}
\begin{equation}
\label{HD trac:1}
\tau^{\hd,\ins}_{jm0} =(2j+1)c_{jm0}-3\brac{\frac{j+1}{j}}^\half c_{jm2}\,
\end{equation}
\begin{equation}
\tau^{\hd,\out}_{jm0} =-(2j+1)c_{jm0}+3\brac{\frac{j}{j+1}}^\half c_{jm2}\,
\end{equation}

\begin{equation}
\tau^{\hd,\out}_{jm2} =3\brac{\frac{j}{j+1}}^\half c_{jm0}-\frac{4+3j+2j^2}{j+1} c_{jm2}\,
\end{equation}
\begin{equation}
\label{HD trac:2}
\tau^{\hd,\ins}_{jm2} =-3\brac{\frac{j+1}{j}}^\half c_{jm0}+\frac{3+j+2j^2}{j} c_{jm2}\,
\end{equation}

\bibliographystyle{unsrt}

\begin{thebibliography}{10}

\bibitem{Alberts}
B.~Alberts, A.~Johnson, J.~Lewis, M.~Raff, K.~Roberts, and P.~Walter.
\newblock {\em Molecular biology of the cell}.
\newblock Garland Publishing Inc., New York, 4th Edition, 2002.

\bibitem{Walde:2010}
P.~Walde.
\newblock Building artificial cells and protocell models: Experimental
  approaches with lipid vesicles.
\newblock {\em Bioessays}, 32:296--303, 2010.

\bibitem{Rumy:2006}
R.~Dimova, S.~Aranda, N.~Bezlyepkina, V.~Nikolov, K.~A. Riske, and R.~Lipowsky.
\newblock A practical guide to giant vesicles. probing the membrane nanoregime
  via optical microscopy.
\newblock {\em J. Phys. Cond. Matt.}, 18:S1151--S1176, 2006.

\bibitem{Abkarian:2008}
M.~Abkarian and A.~Viallat.
\newblock Vesicles and red blood cells in shear flow.
\newblock {\em Soft Matter}, 4:653--657, 2008.

\bibitem{VlahovskaCR}
P.~M. Vlahovska, T.~Podgorski, and M.~Misbah.
\newblock Vesicles and red blood cells: from individual dynamics to rheology.
\newblock {\em Comptes Rendus Physique}, 10:775Ü789, 2009.

\bibitem{softmatter:2009}
R.~Dimova, N.~Bezlyepkina, M.~D. Jordš, R.~L. Knorr, K.~A. Riske, M.~Staykova,
  P.~M. Vlahovska, T.~Yamamoto, P.~Yang, and R.~Lipowsky.
\newblock Vesicles in electric fields: Some novel aspects of membrane behavior.
\newblock {\em Soft Matter}, 5:3201 -- 3212, 2009.

\bibitem{VlahovskaAPLB}
P.~M. Vlahovska.
\newblock Non-equilibrium dynamics of lipid membranes: deformation and
  stability in electric fields.
\newblock In A.~Iglic, editor, {\em Advances in Planar Lipid Bilayers and
  Liposomes, vol. 12}, page in press. Elsevier, 2010.

\bibitem{Kantsler-Steinberg:2005}
V.~Kantsler and V.~Steinberg.
\newblock Orientation and dynamics of a vesicle in tank-treading motion in
  shear flow.
\newblock {\em Phys. Rev. Lett.}, 95:258101, 2005.

\bibitem{Kantsler-Steinberg:2006}
V.~Kantsler and V.~Steinberg.
\newblock Transition to tumbling and two regimes of tumbling motion of a
  vesicle in shear flow.
\newblock {\em Phys. Rev. Lett.}, 96:036001, 2006.

\bibitem{Mader:2006}
M.-A. Mader, V.~Vitkova, M.~Abkarian, A.~Viallat, and T.~Podgorski.
\newblock Dynamics of viscous vesicles in shear flow.
\newblock {\em Eur. Phys. J. E}, 19:389--397, 2006.

\bibitem{Kantsler-Steinberg:2009}
J.~Deschamps, V.~Kantsler, and V.~Steinberg.
\newblock Phase diagram of single vesicle dynamical states in shear flow.
\newblock {\em Phys. Rev. Lett.}, 102, 2009.

\bibitem{Deschamps:2009}
J.~Deschamps, V.~Kantsler, E.~Segre, and V.~Steinberg.
\newblock Dynamics of a vesicle in general flow.
\newblock {\em PNAS}, 106:11444--11447, 2009.

\bibitem{Kummrow-Helfrich:1991}
M.~Kummrow and W.~Helfrich.
\newblock Deformation of giant lipid vesicles by electric fields.
\newblock {\em Phys. Rev. A}, 44:8356--8360, 1991.

\bibitem{Mitov:1993}
M.~D. Mitov, P.~Meleard, M.~Winterhalter, M.~I. Angelova, and P.~Bothorel.
\newblock Electric-field-dependent thermal fluctuations of giant vesicles.
\newblock {\em Phys.\ Rev.\ E}, 48:628--631, 1993.

\bibitem{Riske-Dimova:2005}
K.~A. Riske and R.~Dimova.
\newblock Electro-deformation and poration of giant vesicles viewed with high
  temporal resolution.
\newblock {\em Biophys. J.}, 88:1143--1155, 2005.

\bibitem{Aranda:2008}
S.~Aranda, K.~A. Riske, R.~Lipowsky, and R.~Dimova.
\newblock Morphological transitions of vesicles induced by ac electric fields.
\newblock {\em Biophys. J.}, 95:L19--L21, 2008.

\bibitem{Antonova-Vitkova:2010}
K~Antonova, V~Vitkova, and M.D. Mitov.
\newblock Deformation of giant vesicles in ac electric fields - dependence of
  the prolate-to-oblate transition of vesicles induced by ac electric fields.
\newblock {\em Euro. Phys. Lett.}, 89:38004, 2010.

\bibitem{Riske-Dimova:2006}
K.~A. Riske and R.~Dimova.
\newblock Electric pulses induce cylindrical deformations on giant vesicles in
  salt solutions.
\newblock {\em Biophys. J.}, 91:1778--1786, 2006.

\bibitem{Allan-Mason:1962}
R.~S. Allan and S.~G. Mason.
\newblock Particle behaviour in shear and electric fields. i. deformation and
  burst of fluid drops.
\newblock {\em Proc. Royal Soc. A}, 267:45--61, 1962.

\bibitem{Ha:2000c}
J.~W. Ha and S.~M. Yang.
\newblock Electrohydrodynamic effects on the deformation and orientation of a
  liquid capsule in a linear flow.
\newblock {\em Phys. Fluids}, 12:1671--1684, 2000.

\bibitem{Papageorgiou:2009}
S.~Mahlmann and D.~T. Papageorgiou.
\newblock Numerical study of electric field effects on the deformation of
  two-dimensional liquid drops in simple shear flow at arbitrary reynolds
  number.
\newblock {\em J. Fluid Mech.}, 626:367--393, 2009.

\bibitem{Vlahovska:ER}
P.~M. Vlahovska.
\newblock On the rheology of a dilute emulsion in a uniform electric field.
\newblock {\em J. Fluid. Mech.}, page accepted, 2010.

\bibitem{Vlahovska-Dimova:2009}
P.~M. Vlahovska, R.~S. Gracia, S.~Aranda-Espinoza, and R.~Dimova.
\newblock Electrohydrodynamic model of vesicle deformation in alternating
  electric fields.
\newblock {\em Biophys. J.}, 96:4789--4803, 2009.

\bibitem{Misbah:2006}
C.~Misbah.
\newblock Vacillating breathing and tumbling of vesicles under shear flow.
\newblock {\em Phys. Rev. Lett.}, 96:028104, 2006.

\bibitem{Vlahovska:2007}
P.~M. Vlahovska and R.S. Gracia.
\newblock Dynamics of a viscous vesicle in linear flows.
\newblock {\em Phys. Rev. E}, 75:016313, 2007.

\bibitem{Lebedev:2008}
V.~V. Lebedev, K.~S. Turitsyn, and S.~S. Vergeles.
\newblock Nearly spherical vesicles in an external flow.
\newblock {\em New J. Phys.}, 10:043044, 2008.

\bibitem{Schwalbe:2010}
J.~Schwalbe, P.~M. Vlahovska, and M.~Miksis.
\newblock Monolayer slip effects on the dynamics of a lipid bilayer vesicle in
  a viscous flow.
\newblock {\em J. Fluid Mech.}, 647:403--419, 2010.

\bibitem{Lomholt-Miao:2006}
M.~A. Lomholt and L.~Miao.
\newblock Descriptions of membrane mechanics from microscopic and effective
  two-dimensional perspectives.
\newblock {\em J. Phys. A}, 39:10323--10354, 2006.

\bibitem{Helfrich:1973}
W.~Helfrich.
\newblock Elastic properties of lipid bilayers - theory and possible
  experiments.
\newblock {\em Z. Naturforsch.}, 28c:693--703, 1973.

\bibitem{Canham:1970}
P.~B. Canham.
\newblock The minimum energy of bending as a possible explanation of the
  biconcave shape of the human red blood cell.
\newblock {\em J. Theor. Biol.}, 26:61--81, 1970.

\bibitem{Evans-Skalak}
E.~Evans and R.~Skalak.
\newblock {\em Mechanics and Thermodynamics of Biomembranes}.
\newblock CRC Press, Boca Raton, Florida, 1980.

\bibitem{Seifert:1997}
U.~Seifert.
\newblock Configurations of fluid membranes and vesicles.
\newblock {\em Advances in physics}, 46:13--137, 1997.

\bibitem{Melcher-Taylor:1969}
J.~R. Melcher and G.~I. Taylor.
\newblock Electrohydrodynamics - a review of role of interfacial shear stress.
\newblock {\em Annu. Rev. Fluid Mech.}, 1:111--146, 1969.

\bibitem{Saville:1997}
D.~A. Saville.
\newblock Electrohydrodynamics: The taylor-melcher leaky dielectric model.
\newblock {\em Annu. Rev.Fluid Mech.}, 29:27--64, 1997.

\bibitem{Taylor:1966}
G.~I. Taylor.
\newblock Studies in electrohydrodynamics. i. circulation produced in a drop by
  an electric field.
\newblock {\em Proc. Royal Soc. A}, 291:159--166, 1966.

\bibitem{JonesTB}
T.~B. Jones.
\newblock {\em Electromechanics of particles}.
\newblock Cambridge University Press, New York, 1995.

\bibitem{Grosse-Schwan:1992}
C.~Grosse and H.~P. Schwan.
\newblock Cellular membrane potentials induced by alternating fields.
\newblock {\em Biophys. J.}, 63:1632--1642, 1992.

\bibitem{Schwan}
H.~P. Schwan.
\newblock Dielectrophoresis and rotation of cells.
\newblock In E.~Neumann, A.~E. Sowers, and C.~A. Jordan, editors, {\em
  Electroporation and electrofusion in cell biology}, pages 3--21. Plenum
  Press, 1989.

\bibitem{Kinosita:1988}
K.~Kinosita~Jr., I.~Ashikawa, N.~Saita, H.~Yoshimura, H.~Itoh, K.~Nagayama, and
  A.~Ikegami.
\newblock Electroporation of cell membrane visualized under a pulsed laser
  fluorescence microscope.
\newblock {\em Biophys. J.}, 53:1015--1019, 1988.

\bibitem{Seifert:1999}
U.~Seifert.
\newblock Fluid membranes in hydrodynamic flow fields: Formalism and an
  application to fluctuating quasispherical vesicles.
\newblock {\em Eur. Phys. J. B}, 8:405--415, 1999.

\bibitem{Dimova-Aranda:2007}
R.~Dimova, K.~A. Riske, S.~Aranda, N.~Bezlyepkina, R.~L. Knorr, and
  R.~Lipowsky.
\newblock Giant vesicles in electric fields.
\newblock {\em Soft matter}, 3:817--827, 2007.

\bibitem{Portet:2009}
T.~Portet, F.C.I. Febrer, J.M. Escoffre, C.~Favard, M.P. Rols, and D.~S. Dean.
\newblock Visualization of membrane loss during the shrinkage of giant vesicles
  under electropulsation.
\newblock {\em Biophys. J.}, 96:4109--4121, 2009.

\bibitem{Riske-Dimova:2009}
K.~A. Riske, R.~L. Knorr, and R.~Dimova.
\newblock Bursting of charged multicomponent vesicles subjected to electric
  pulses.
\newblock {\em Soft Matter}, 5:1983--1986, 2009.

\bibitem{Gracia:2010}
R.~S. Gracia, N.~Bezlyepkina, R.~L. Knorr, R.~L. Lipowsky, and R.~Dimova.
\newblock Effect of cholesterol on the rigidity of saturated and unsaturated
  membranes: fluctuation and electrodeformation analysis of giant vesicles.
\newblock {\em Soft Matter}, 6:1472 -- 1482, 2010.

\bibitem{Portet:2010}
T.~Portet and R.~Dimova.
\newblock A new method for measuring edge tensions in lipid membranes: Effect
  of membrane composition.
\newblock {\em submitted}.

\bibitem{Knorr:2010}
R.~L. Knorr, M.~Staykova, R.~S. Gracia, and R.~Dimova.
\newblock Wrinkling and electroporation of giant vesicles in the gel phase.
\newblock {\em Soft Matter}, page DOI:10.1039/b925929e, 2010.

\bibitem{Needham-Hochmuth:1989}
D.~Needham and R.~M. Hochmuth.
\newblock Electromechanical permeabilization of lipid vesicles. role of
  membrane tension and compressibility.
\newblock {\em Biophys. J.}, 55:1001--1009, 1989.

\bibitem{Kantsler-Segre-Steinberg:2008b}
V.~Kantsler, E.~Segre, and V.~Steinberg.
\newblock Dynamics of interacting vesicles and rheology of vesicle suspension
  in shear flow.
\newblock {\em Europhys. Lett.}, 82:58005, 2008.

\bibitem{Vitkova:2008}
V.~Vitkova, M.~Mader, B.~Polack, C.~Misbah, and T.~Podgorski.
\newblock Micro-macro link in rheology of erythrocyte and vesicle suspensions.
\newblock {\em Biophys. J.}, {95}({6}):{L33--L35}, {2008}.

\bibitem{Coupier:2008}
G.~Coupier, B.~Kaoui, T.~Podgorski, and C.~Misbah.
\newblock {Noninertial lateral migration of vesicles in bounded Poiseuille
  flow}.
\newblock {\em {Phys. Fluids}}, {20}({11}), {NOV} {2008}.

\bibitem{Callens:2008}
N.~Callens, C.~Minetti, G.~Coupier, M.-A. Mader, F.~Dubois, C.~Misbah, and
  T.~Podgorski.
\newblock Hydrodynamic lift of vesicles under shear flow in microgravity.
\newblock {\em Europhys. Lett.}, 83:24002, 2008.

\bibitem{Schwalbe:thesis}
J.~Schwalbe.
\newblock {\em Dynamics and stability of lipid bilayer membranes in viscous
  flow and electric fields}.
\newblock PhD thesis, Northwestern University, 2010.

\bibitem{Keller-Skalak:1982}
S.~R. Keller and R.~Skalak.
\newblock Motion of a tank -reading ellipsoidal particle in shear flow.
\newblock {\em J. Fluid Mech.}, 120:27--47, 1982.

\bibitem{Evans-Rawicz:1990}
E.~Evans and W.~Rawicz.
\newblock Entropy driven tension and bending elasticity in condensed-fluid
  membranes.
\newblock {\em Phys.\ Rev.\ Lett.}, 64:2094--2097, 1990.

\bibitem{Blawzdziewicz-Vlahovska-Loewenberg:2000}
J.~B{\l}awzdziewicz, P.~Vlahovska, and M.~Loewenberg.
\newblock Rheology of a dilute emulsion of surfactant-covered spherical drops.
\newblock {\em Physica A}, 276:50--80, 2000.

\bibitem{Vlahovska:2005}
P.~Vlahovska, J.~B{\l}awzdziewicz, and M.~Loewenberg.
\newblock Deformation of a surfactant-covered drop in a linear flow.
\newblock {\em Phys.\ Fluids}, 17:Art. No.103103, 2005.

\bibitem{Danker:2007b}
G.~Danker, T.~Biben, T.~Podgorski, C.~Verdier, and C.~Misbah.
\newblock Dynamics and rheology of a dilute suspension of vesicles: higher
  order theory.
\newblock {\em Phys. Rev. E}, 76:041905, 2007.

\bibitem{Vlahovska:2003}
P.M. Vlahovska.
\newblock {\em Dynamics of a surfactant-covered drop and the non-Newtonian
  rheology of emulsions}.
\newblock PhD thesis, Yale University, 2003.
\newblock pdf file available by email: petia@aya.yale.edu.

\bibitem{Vlahovska:2009a}
P.~Vlahovska, J.~B{\l}awzdziewicz, and M.~Loewenberg.
\newblock Small-deformation theory for a surfactant-covered drop in linear
  flows.
\newblock {\em J. Fluid Mech.}, 624:293--337, 2009.

\bibitem{Cichocki-Felderhof-Schmitz:1988}
B.~Cichocki, B.~U. Felderhof, and R.~Schmitz.
\newblock Hydrodynamic interactions between two spherical particles.
\newblock {\em PhysicoChem. Hyd.}, 10:383--403, 1988.

\end{thebibliography}

\end{document}